\documentclass[runningheads]{llncs}
\usepackage{graphicx}
\usepackage{amsmath}
\usepackage{amsfonts}
\usepackage{float}
\usepackage{import}
\usepackage{longtable}
\usepackage{mathtools}
\allowdisplaybreaks
\usepackage{color}

\begin{document}
\title{
	Automated Deep Abstractions for \\ Stochastic Chemical Reaction Networks
	\thanks{
		TP's research is supported by the Ministry of Science, Research and the Arts of the state of Baden-W\"{u}rttemberg, 
		and the DFG Centre of Excellence 2117 `Centre for the Advanced Study of Collective Behaviour' (ID: 422037984),
		DR's research is supported by Young Scholar Fund (YSF), project no. $P83943018 FP 430\_/18$ and by the `Centre for the Advanced Study of Collective Behaviour'. 
The authors would like to thank to Luca Bortolussi for inspiring discussions on the topic.
	}
}
\titlerunning{Automated Deep Abstractions}
\author{
	Denis Repin%\inst{1}\orcidID{???}
	\and
	Tatjana Petrov%\inst{1}\orcidID{???}
}
\authorrunning{D. Repin, T.Petrov}
\institute{
	{Department of Computer and Information Sciences, University of Konstanz, Konstanz, Germany}\\
	\and
	{Centre for the Advanced Study of Collective Behaviour, University of Konstanz, 78464, Konstanz, Germany}
%\and
%	{Max-Planck-Institut für Ornithologie, Am Obstberg 1, 78315 Radolfzell, Germany}
}
\maketitle
\begin{abstract}
Predicting stochastic cellular dynamics as emerging from the mechanistic models of molecular interactions  is a long-standing challenge in systems biology:
low-level chemical reaction network (CRN) models give raise to a highly-dimensional continuous-time Markov chain (CTMC) which is computationally demanding and often prohibitive to analyse in practice. 
A recently proposed abstraction method uses deep learning to replace this CTMC with a discrete-time continuous-space process, 
by training a mixture density deep neural network with traces sampled at regular time intervals (which can obtained either by simulating a given CRN or as time-series data from experiment). 
The major advantage of such abstraction is that it produces a computational model
that is dramatically cheaper to execute, while preserving the statistical features of the training data. 
In general, the abstraction accuracy improves with the amount of training data.
However, depending on a CRN, the overall quality of the method -- the efficiency gain and abstraction accuracy --  will also depend on the choice of neural network architecture given by hyper-parameters such as the layer types and connections between them. 
As a consequence, in practice, the modeller would have to take care of finding the suitable architecture manually, for each given CRN, through a tedious and time-consuming trial-and-error cycle.  

In this paper, we propose to further automatise deep abstractions for stochastic CRNs, 
through learning the optimal neural network architecture along with learning the transition kernel of the abstract process. 
Automated search of the architecture makes the method applicable directly to any given CRN, 
which is time-saving for deep learning experts and crucial for non-specialists. 
We implement the method and demonstrate its performance on a number of representative CRNs with multi-modal emergent phenotypes. 
\end{abstract}
\section{Introduction}
Understanding how the dynamics of complex biological systems such as 
a biological cell or a group of interacting cells 
emerges from the dynamics of low-level molecular interactions 
is one of the key challenges of systems and synthetic biology.
%
%could inform disease treatment and drug discovery. 
%
Low-level mechanisms of molecular interactions are usually hypothesised in form of chemical reaction networks.
%For example, the following set of reactions constitutes a reaction network with three species $S, I, R$, and three reactions:
%%\vspace*{-35pt}
%%\begin{center}
%\begin{align*}%{c}
%	S \rightarrow I , \;\;\;
%	S + I \rightarrow 2 I , \;\;\;
%	I \rightarrow R 
%\end{align*}
%%\end{center}
%%\vspace*{-35pt}
%The model above is the well-known epidemiological SIR model that describes the spread of an infectious disease, where
%%\begin{itemize}
%%	\item[$\bullet$] 
%	susceptible (\emph{S}) species may become infected (\emph{I}),
%%	\item[$\bullet$] 
%	infectious species can spread the disease by infecting susceptible species, 
%%	\item[$\bullet$] 
%	or get recovered (\emph{R}). 
%%\end{itemize}
Each reaction fires with a corresponding rate. 
A reaction network induces a stochastic dynamical system - continuous-time Markov chain (CTMC), describing how the state - a vector ${{\eta}_t}$ enumerating multiplicities of each of the species - changes over time upon firing of reactions.
% of events. 
Computationally predicting the distribution of species over time from such CTMC is generally challenging, 
due to a huge number of reachable states, due to stochasticity and events happening at multiple time-scales.
%The challenge becomes even more prominent if we aim to predict how and which system dynamics can emerge at larger organisational scales.
%
%in multi-scale modelling, where we aim to explain phenomena happening at the level of a collective (e.g. tissues), emerging as a product of internal and communication processes of many individuals (e.g. cells) living in a shared environment. 
%Moreover,
%%
%Instead, we wish to predict the system state at some interesting time intervals, hence abstracting away all intermediate states and reactions, while preserving the statistical properties of the original model executed at low level granularity. 
%% through a higher-level model. 
%% 
%Model abstraction aims at constructing simpler models that are easier to analyse, 
%%, allowing to generate approximate trajectories in a significantly faster way than the original detailed model, 
%while retaining the most important of the original model of interest. %a reasonable accuracy.
%
Two major approaches are used to analyse the CTMC. 
The first approach focuses on computing the transient evolution of the probability related to each state of the CTMC numerically. %, referred to as the transient distribution. 
The transient distribution evolves according to the Kolmogorov forward equation (chemical master equation in the chemistry literature), and it is typically very difficult to solve the forward equations except for the simplest systems.
%, sophisticated numerical algorithms have been designed to numerically solve the forward equation for larger systems \cite{}.
The second approach is based on a statistical estimation of trace distribution and event probabilities of the CTMC by generating many sample traces \cite{Gillespie}, often referred to as stochastic simulation Gillespie algorithm (SSA).
While this method generally allows to trade-off computational tractability with loosing precision, % of the transient distribution,  
even simulating a single trace can still take considerable processor time, 
%due to a large number of reachable states, due to stochasticity and
especially when some reactions fire at very fast time-scales relative to the global time horizon of interest. % duration.
At the same time, we are often not interested in predictions at such small time-scales or transient distributions for each of the species. % (reaction products).
For all these reasons, it is desirable to develop %of major importance to provide 
model reduction techniques for stochastic reaction networks, which allow for efficient simulation, yet faithfully abstract the context-relevant emerging features of the hypothesised mechanism.

\begin{figure}[t]
	\centering
	\includegraphics[width=0.9\linewidth, height=0.4\linewidth, trim=10 10 10 10, clip]{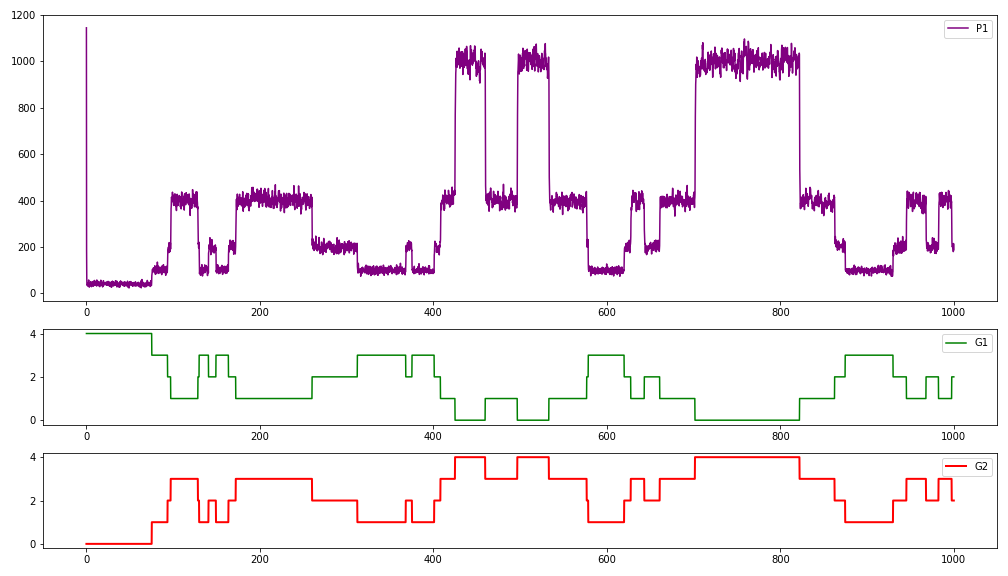}
	\vspace{-5pt}\caption{\footnotesize Sample trajectory of Ex.(\ref{eq:X16_}) network.}
\label{fig:X16_}
\end{figure}

\begin{figure}[t]
	\centering
	\includegraphics[width=0.23\linewidth, trim=30 10 0 20, clip]{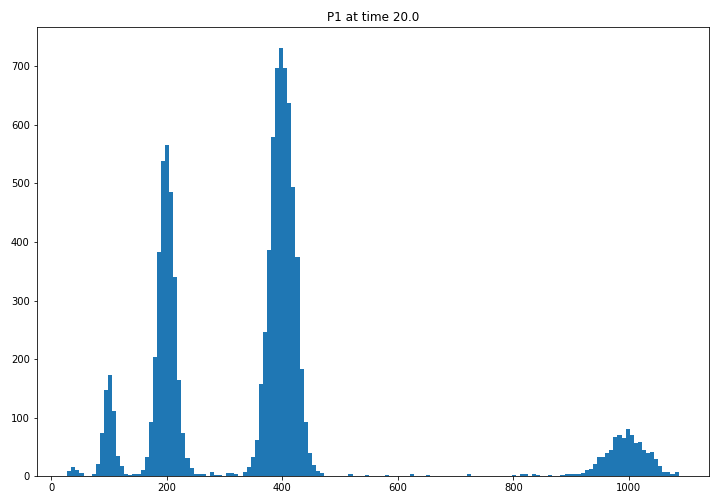}
	\includegraphics[width=0.23\linewidth, trim=30 10 0 20, clip]{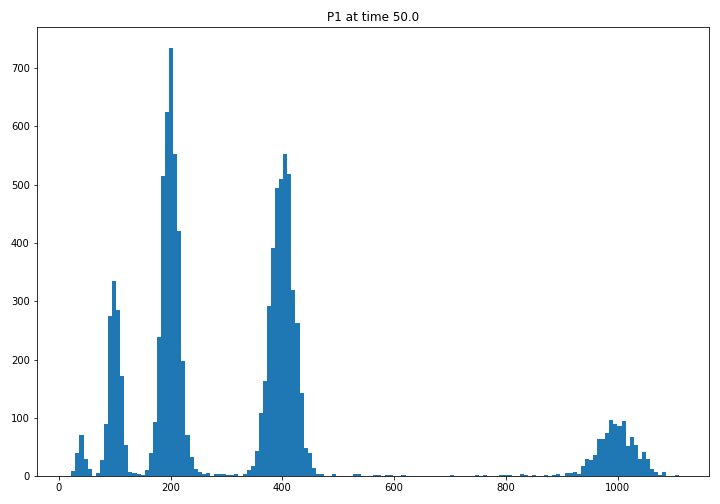}
	\includegraphics[width=0.23\linewidth, trim=30 10 0 20, clip]{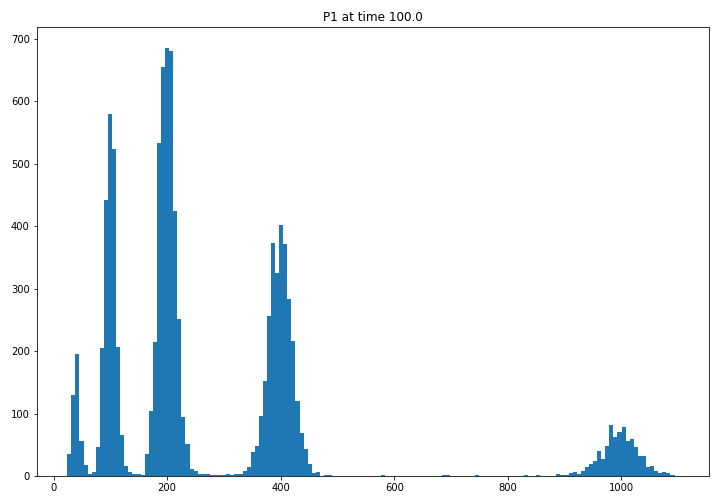}
	\includegraphics[width=0.23\linewidth, trim=30 10 0 20, clip]{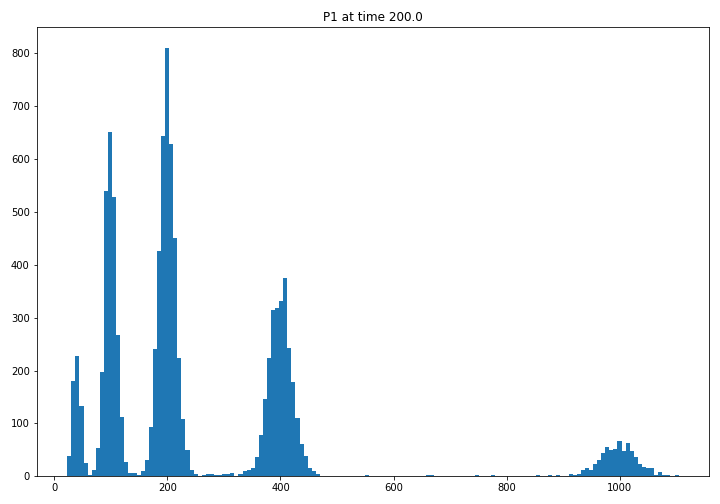}
	\vspace{-5pt}\caption{\footnotesize Distribution (histogram) of the protein $P_1$ at times 20, 50, 100, and 200 for Ex.(\ref{eq:X16_}) network.}
	\label{fig:X16_hist}
\end{figure}

\begin{example} [running example] %\rm
For example, the following set of reactions constitutes a reaction network with three species $G_1, G_2, P_1$, and six reactions:
\begin{align}
%\begin{split}
%& 
G_1   \xrightleftharpoons[\alpha_{12}]{\alpha_{11}}                     G_1 + P_1 , \;
%& 
G_2   \xrightarrow{\alpha_{21}}                                         G_2 + P_1,  \;
%& 
P_1   \xrightarrow{\beta_{1}}                                           \emptyset,  \;
%& 
G_1   \xrightleftharpoons[\epsilon \gamma_{21}]{\epsilon \gamma_{12}}   G_2,
%\end{split}
\label{eq:X16_}
\end{align}
where $0 < \epsilon \ll 1$. 
This fast-and-slow network (\cite{crn_models}, Eq. 16) may be interpreted as a gene slowly switching between two different expression modes.
In Fig.\ref{fig:X16_}, we show a sample trajectory for Ex.(\ref{eq:X16_}), where one can see notable differences in species' abundance and reaction time-scales.
Moreover, we may be interested only in abstractions reproducing the distribution of protein $P_1$ at several interesting time points, e.g. four at time points shown in Fig.\ref{fig:X16_hist}. 
%, we show how   evolves in time. 
%
\end{example}

Deep abstractions, introduced in \cite{bortolussi_abstraction}, propose to use available simulation algorithms to generate a suitable number of simulated system traces, and then learn an abstract model from such data. 
The task of learning a transition kernel for a Markov process that defines the abstract model is solved as a supervised learning problem:
the transition kernel for this Markov process is modelled as a probability mixture with parameters depending on the system state, and a deep neural network is trained on simulated data to parameterise this probability mixture.
Such abstract  model preserves the statistics of the original network dynamics, but runs on a discrete time-scale representing equally distributed time intervals, abstracting away all intermediate transitions, which can lead to significant computational savings. 

\subsubsection{Contributions.}
The performance of any deep learning application largely depends on the choice of the neural network architecture, usually constructed by the user through a trial-and-error process. 
In context of applying deep abstractions proposed in  \cite{bortolussi_abstraction}, this means that the modeller would have to take care of finding the suitable architecture manually, for each given CRN.
The main contribution of this paper is a framework for deep abstractions where the neural network architecture search is automated: in parallel to learning the kernel of the stochastic process, we learn a neural network architecture, by employing the recent advances on this topic in the deep learning community \cite{darts,proxyless_nas}.
We implement our technique as a Python library StochNetV2 available on GitHub and we illustrate the quality of model reduction on different reaction network case studies.

\subsubsection{Related Works.}

Different techniques on reducing stochastic CRNs have been proposed in literature and practice over the years. 
Classical limit approximations of deterministic limit, moments or mean-field approximation  \cite{Anderson2010,cardelli2016stochastic}
can provide significant computational savings, 
but they do not apply to general stochastic CRNs, 
especially when species distributions emerging over time are non-Gaussian, as for example is the case shown in Ex.(\ref{eq:X16_}). 
%
%Then, a stochastic description of chemical reactions is mandatory to analyse the behaviour of the system. 
Moreover, principled model reduction techniques have been proposed in several %bisimulation, lumpability 
aggregation \cite{cardelli2017syntactic,approx,journal,mathpaper,lumpability,tribastone2018speeding,HC1} and time-scale separation frameworks \cite{verenaHybrid,hsb2015,gunawardena2012linear,pahlajani2011stochastic}.
These techniques are generally based on detecting species, reactions or states which are behaviourally indistinguishable or similar. 
In these methods, the space of considered abstractions is typically discrete and as a consequence, it does not allow smooth tuning of abstracted entities, or control of abstraction accuracy. 
In other words, the achieved abstraction may or may not be significant, and once the method is applied, it is difficult to further improve it, both in terms of abstraction size, and accuracy. 
This is different to deep abstractions, where the abstraction accuracy can be improved by increasing the model size and/or adding more training data, and increasing time discretisation interval improves abstraction efficiency.
Abstractions based on statistical analysis of traces, 
closest to the idea of deep abstractions in \cite{bortolussi_abstraction}, 
include 
\cite{palaniappan}, who proposed to construct abstractions using information theory to discretise the state space and select a subset of all original variables and their mutual dependencies and a Dynamic Bayesian Network is constructed to produce state transitions, 
as well as a statistical approach to approximate dynamics of fast-and-slow models was developed by %Michaelides et al.
 in \cite{multiscale_abstraction}, where Gaussian Processes are used to predict the state of the fast equilibrating internal process as a function of the environmental state.
It is worth noting that all the mentioned reduction techniques, except from the exact frameworks based on syntactic criteria, such as in \cite{mathpaper,cardelli2017syntactic}, do not guarantee error bounds a priori. 

%Here we present an example model which will be used across the paper for demonstration purposes.
%Consider the following fast-and-slow network (\cite{crn_models}, Eq. 16) displays interesting dynamics with multimodal distribution of the protein, see Fig.\ref{fig:X16_hist}.
%%
%\begin{equation}
%\begin{split}
%& G_1   \xrightleftharpoons[\alpha_{12}]{\alpha_{11}}                     G_1 + P_1  \\
%& G_2   \xrightarrow{\alpha_{21}}                                         G_2 + P_1  \\
%& P_1   \xrightarrow{\beta_{1}}                                           \emptyset  \\
%& G_1   \xrightleftharpoons[\epsilon \gamma_{21}]{\epsilon \gamma_{12}}   G_2, \ \ 0 < \epsilon \ll 1
%\end{split}
%\label{eq:X16_}
%\end{equation}
%%
%Network (\ref{eq:X16_}) may be interpreted as describing a gene slowly switching between two expressions $\textrm{G}_1$ and $\textrm{G}_2$. 
%When in state $\textrm{G}_1$, the gene produces and degrades protein $\textrm{P}_1$, while when in state $\textrm{G}_1$, it only produces $\textrm{P}_1$, but generally at a different rate than when it is in state $\textrm{G}_1$. 
%Furthermore, $\textrm{P}_1$ may also spontaneously degrade.

%\begin{figure}[t]
%	\centering
%	\includegraphics[width=0.425\linewidth]{images/X16/X16_hist_0_time20.png}%
%	\includegraphics[width=0.425\linewidth]{images/X16/X16_hist_0_time50.png}
%	\includegraphics[width=0.425\linewidth]{images/X16/X16_hist_0_time100.png}%
%	\includegraphics[width=0.425\linewidth]{images/X16/X16_hist_0_time200.png}
%	\caption{\footnotesize Distribution (histogram) of the protein $P_1$ at times 20, 50, 100, and 200.}
%	\label{fig:X16_hist}
%\end{figure}

%
%\end{example}
\section{Backgound and Preliminaries}

\subsubsection{Neural Networks.}
In the last decade, deep neural networks (DNNs, NNs) gathered a lot of attention from researchers as well as from industry, bringing breakthroughs in various application areas, such as computer vision, time-series analysis, speech recognition, machine translation, etc.
Neural networks are well known as a powerful and versatile framework for high-dimensional learning tasks.
The key feature of neural networks is that they can represent an arbitrary function, mapping a set of input variables $( x_1,\dots,x_n )$ to a set of output variables $(y_1,\dots,y_m)$. Further we denote them as $x$ and $y$ for simplicity.

Neural networks are typically formed by composing together many different functions.
The model is associated with a directed acyclic graph describing how the functions are composed together. 
For example, we might have three functions $f_1, f_2, f_3$ connected in a chain to form $f(x) = f_3(f_2(f_1(x)))$.
In this case, $f_1$ is called the first layer of the network, $f_2$ is called the second layer, and so on. 
The outputs of each layer are called {\it features}, or {\it latent representation} of the data.
By adding more layers and more units within a layer, a deep network can represent functions of increasing complexity \cite{deeplearningbook}.

\begin{figure}[]
		\centering
		\includegraphics[width=0.45\linewidth]{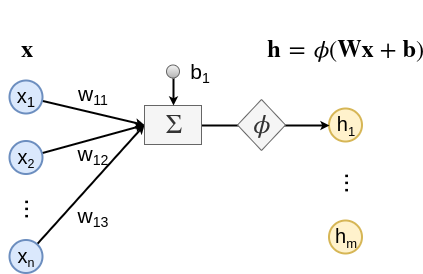}
		\caption{\footnotesize A single layer of a neural network. Outputs are computed as a linear transformation ${\bf Wx + b}$ followed by a non-linear activation function $\phi$}
		\label{fig:neural_net}
\end{figure}

In particular, each layer usually computes a linear transformation ${W x + b}$ where $W$ is a weight matrix and $b$ is bias vector. Additional nonlinearities are inserted in between the layers which allow the network to represent arbitrary nonlinear functions (see the illustration in Fig.\ref{fig:neural_net}.

NNs are trained on a set of training examples $ \{(x, y)\}$ with the aim not to memorize the data, but rather to learn the underlying dependencies within the variables, so that the output $y$ can be predicted for unseen values of $x$.
During training, the weights in a network are adjusted to minimize the learning objective - loss function - on training data.
The quality of a trained model is usually measured on unseen (test) data, which is addressed as the model's generalization ability.
\subsubsection{Mixture Density Networks.}
However, conventional neural networks perform poorly on a class of problems involving the prediction of continuous variables, for which the target data is multi-valued.
Minimising a sum-of-squares error encourages a network to approximate the conditional average of the target data, which does not capture the information on the distribution of output values. % \cite{mdn}.
\begin{figure}[t]
	\centering
	\includegraphics[width=0.85\linewidth]{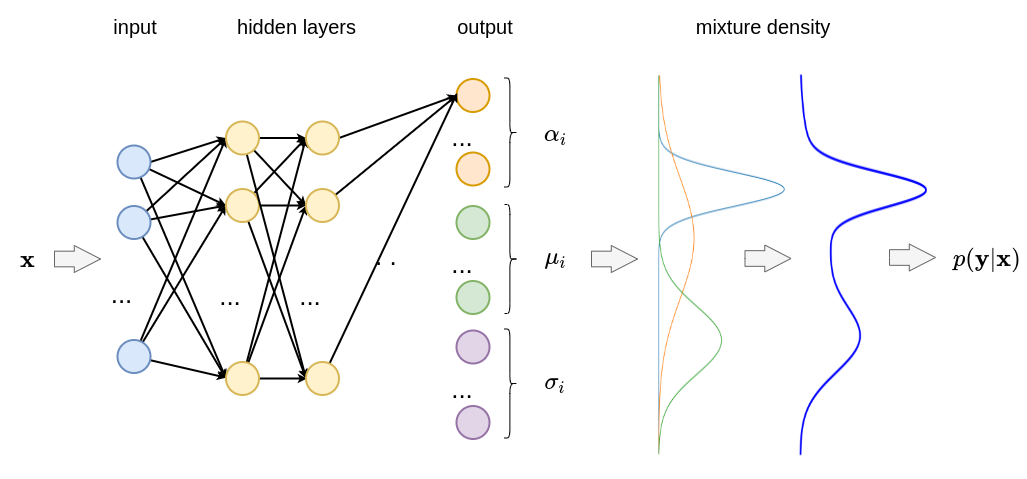}
	\caption{\footnotesize Mixture Density Network structure. Given ${\bf x}$, neural network outputs values $ \mu_i $ and $ \sigma_i , i=1,...,m$ that define $m$ Gaussian distributions. Weighted by mixing coefficients $\alpha_i$, they form a mixture density - conditional probability density $p({\bf y}|{\bf x})$.}
	\label{fig:mdn}
\end{figure}

Mixture Density Networks (MDN), proposed in \cite{mdn}, is a class of models which overcome these limitations by combining a conventional neural network with a mixture density model, illustrated in Fig \ref{fig:mdn}.
This neural network provides the parameters for multiple distributions, which are then mixed by the weights (also provided by the network).
Therefore Mixture Density Networks can in principle represent arbitrary conditional distributions, in the same way that a conventional neural network can represent arbitrary non-linear functions. 

To construct an abstract model defined by a Markov process, we need to learn its transition kernel. 
In other words, given a vector $\eta_{t}$ describing the system state at time $t$, we wish to predict the state $\eta_{t + \Delta t}$, which follows a distribution, conditioned on $\eta_{t}$. Therefore we can use Mixture Density Networks to learn the conditional distribution $p(\eta_{t + \Delta t} | \eta_{t})$.
\section{Deep Abstractions}
%
%For further details about model abstraction procedure we refer to the original paper \cite{bortolussi_abstraction}.
%TODO: model anstraction, simulation with abstracted model, and histogram distance as a measure of quality
%
Here we present the main steps of the abstraction technique originally proposed 
%by L. Bortolussi. For more details we refer to the original paper 
in \cite{bortolussi_abstraction}.
For more details we refer to the original paper.
\subsubsection{Abstract Model.}
Let $\{ \eta_{t} \}_{t\ge0}$ be the CTMC describing a CRN with the state space $S = \mathbb{N}^m$.
As mentioned earlier, with the abstract model we wish to reproduce the dynamics of the original process at a fixed temporal resolution. 
%For this we fix a time step $\Delta t$ and an initial time instant $t_0$ and define
Let $\{ \tilde{\eta}_{i} \}_{i\in \mathbb{N}}$ be a discrete-time stochastic process such that 
\begin{equation}
	\tilde{\eta}_{i} := \eta_{t_0 + i \Delta t} \ \ \ \ \forall i \in \mathbb{N}
\end{equation}
\noindent with fixed time interval $\Delta t$ and initial time $t_0$.
The resulting process $\tilde{\eta}_{i}$ is a time-homogeneous discrete-time Markov chain (DTMC), with a transition kernel 
\begin{equation}
	K_d(s, s_0) = \mathbb{P}(\eta_{\Delta t} = s \ | \ \eta_{0} = s_0) \ \ \ \ \forall s_0, s \in S.
\end{equation}

\noindent Further, two following approximations take place:
\begin{itemize}
	\item[1.] The state space $S = \mathbb{N}^m$ is embedded into the continuous space $\tilde{X} = \mathbb{R}_{\ge 0}^m$. The abstract model takes values in $\tilde{X}$.
	\item[2.] The kernel $K_d$ is approximated by a new kernel $K(x | x_0)$ operating in the continuous space $\tilde{X}$. The kernel $K(x | x_0)$ is modelled by a MDN.
\end{itemize}

\noindent To evaluate the abstract model, we introduce a time-bounded \emph{reward function} $r$ that monitors the properties we wish to preserve in the reduced model. 
This function, therefore, maps from the space of discrete-time trajectories $S^M$ to an arbitrary space $T$ (here $M$ is an upper bound on the length of discrete-time trajectories, and $\tilde{\eta}_{[0, M]}$ denotes time-bounded trajectories).
For example, it can be a projection, counting the populations of a subset of species, or it can take Boolean values corresponding to some linear temporal logic properties.
Note that $r(\tilde{\eta}_{[0, M]})$ is a probability distribution on $T$.

As an error metric we use the distance $d$ between the abstract distribution and $r(\tilde{\eta}_{[0, M]})$, which is evaluated statistically, as the distance among histograms,~$h$~\cite{acc_measurment}. In our experiments as a distance $d$ we use $L_1$ metric:  %\tanja{between distributions? what are X and Y?}\dn{updated this text, should be better (X and Y are distributions)}:
%As an error metric we use the distance between the abstract distribution and $r(\tilde{\eta}_{[0, M]})$, which is evaluated statistically, as the distance $d$ among histograms \cite{acc_measurment}. In our experiments we use $L_1$ distance \tanja{between distributions? what are X and Y?}\tanja{in the end, for space savers, we may avoid indenting these two equations}:
\begin{equation}
	d(X, Y) = \sum_z |h_X(z) - h_Y(z)|,
	\label{eq:l1_dist}
\end{equation}
\noindent or Intersection over Union (IoU) distance:
\begin{equation}
	d(X, Y) = \frac{\sum_z {\rm min}(h_X(z), h_Y(z))}{\sum_z {\rm max}(h_X(z), h_Y(z))}.
	\label{eq:iou_dist}
\end{equation}

\noindent Here is the formal definition of model abstraction \cite{bortolussi_abstraction}:
\begin{definition} \rm
	Let $\{\eta_i\}_{i=0}^M$ be a discrete time stochastic process over an arbitrary state space $S$, with $M \in \mathbb{N}_+$ a time horizon, and let $r \ : \ S^M \rightarrow T$ be the associated reward function. An abstraction of $(\eta, r)$ is a tuple $(\bar{S}, p, \bar{r}, \bar{\eta} = \{\bar{\eta}_i\}_{i=0}^M)$ where:
	\begin{itemize}
		\item $\bar{S}$ is the abstract state space;
		\item $p \ : \ S \rightarrow \bar{S}$ is the abstraction function;
		\item $\bar{r} \ : \ \bar{S}^M \rightarrow T$ is the abstract reward;
		\item $\bar{\eta} = \{\bar{\eta}_i\}_{i=0}^M$ is the abstract discrete time stochastic process over $\bar{S}$.
	\end{itemize}
	\noindent Let $\epsilon > 0$. $\bar{\eta}$ is said to be $\epsilon$-close to $\eta$ with respect to $d$ if, for almost any $s_0 \in S$,
	\begin{equation}
		d(r(\tilde{\eta}_{[0, M]}), \bar{r}(\bar{\eta}_{[0, M]})) < \epsilon \ \ \ \ \ \hbox{conditioned on }\eta_0 = s_0, \ \bar{\eta}_0 = p(s_0)
		\label{eq:e-close}
	\end{equation}
\end{definition}

The simplest choice for the abstraction function could be an identity mapping. Alternatively, one can follow \cite{palaniappan} to identify a subset of species having the most influence on the reward function. 
Inequality (\ref{eq:e-close}) is typically experimentally verified simulating a sufficiently high number of trajectories from both the original system $\tilde{\eta}$ and the abstraction $\bar{\eta}$ starting from a common initial setting.
As there is no way to ensure that the inequality holds for almost any $s_0$ in $S$, we evaluate it for many different initial settings that the model did not see during training. 
Evaluation examples are presented in supplementary material.

\begin{example} [running example cont'd] %\rm
The abstract model for our example shown in (\ref{eq:X16_}) as follows:
\begin{itemize}
	\item The abstract state space $\bar{S}$ is $\mathbb{R}_{\ge 0}^3$, i.e. the continuous approximation of $S=\mathbb{N}^3$;
	\item The abstraction function $p$ is the identity function that maps each point of $S$ into its continuous embedding in $\bar{S}$;
	\item The reward function $r$ is the projection on the protein $P_1$;
	\item The discrete time stochastic process $\bar{\eta}=\{\tilde{\eta}\}_{i=0}^M$ is a DTMC with transition kernel represented by an MDN trained on simulation data.
\end{itemize}
\end{example}
\subsubsection{Dataset Generation.} 
We build our datasets as a sets of pairs $\mathcal{D}:=\{(x, y)\}$ where each $y$ is a sample from the distribution $\mathbb{P}(\eta_{t_0 + \Delta t} \ | \ \eta_{t_0} = x)$, i.e. each pair corresponds to a single transition in discrete-time trajectories $\tilde{\eta}$.
%\tanja{be uniform with sticking to a superscript  with brackets where I previously suggested - at the sentence 'NNs are trained on a set of training examples ...'}
For this, we simulate trajectories starting from random initial settings from $t_0$ to $t_0 + M\Delta t$, and take the consecutive states $(\eta_{t_0 + i\Delta t}, \eta_{t_0 + (i+1)\Delta t}),\ i\in \{0,...,M-1\}$ as $(x, y)$ pairs
%\tanja{is it $k$ or $M$ the horizon?}\dn{M is time horizon for evaluation, so they do not have to be the same in general: for histogram dataset $k \ge M$ as a must. For clarity we can write M here}.

%\tanja{the next paragraph may go in the discussion, maybe - not sure how it could fit}\dn{Yes, it would better fit to the CRN Models and Simulation Data subsection in Implementation section than here.}
%To increase generalisation capabilities of the model, it is important to build the dataset that covers the most variability of the original process.
%Although having more training data is always beneficial for a model, it increases training time.
%Therefore, depending on the variation of trajectories starting from the same initial conditions, we might prefer to run a few simulations for many initial conditions or more simulations for fewer initial conditions.
%When generating the dataset for evaluation, to make histograms more consistent, we usually simulate much more trajectories (from 1000 to 10000) for several initial settings.

\begin{example} [running example: dataset] %\rm
For the example network \ref{eq:X16_}, we simulate 100 trajectories for each of 100 random initial settings. We run simulations up to 200 time units, and fix the time step $\Delta t$ to 20 time units for both training and evaluation (histogram) dataset. Therefore the time horizon $M$ for evaluation is 10 steps.
\end{example}
\subsubsection{Model Training.}
Let $\mathcal{M}$ be a parameterized family of mixture distributions and $g_{\theta}$ be an MDN, where $\theta$ are network weights.
Then, for every input vector $x$, $g_{\theta}(x) \in \mathcal{M}$. 
During training, weights $\theta$ are optimized to maximize the likelihood of samples $y$
%\tanja{consistency wrt. notation for samples - was it bold or with superscipt and brackets?}\dn{it's like this now} 
w.r.t. the parameterized distribution $g_{\theta}(x)$, so that the kernel $K_d$ is approximated by $K$:

\begin{equation}
	K_d(s \ | \ s_0) = \mathbb{P}(\eta_{\Delta t} = s \ | \ \eta_0 = s_0) \approx \mathbb{P}(g_{\theta}(s_0) \in B_s) := K(B_s \ | \ s_0)
\end{equation}
\noindent where $B_s := \{ x \in \tilde{X} \ | \  \|x - s\|  < \frac{1}{2} \}$ is the infinity norm ball with radius $\frac{1}{2}$ centered in $s$, needed to properly compare approximating continuous distribution with the original discrete distribution.
Though model training is a relatively time-consuming task, once we have a trained model, sampling states is extremely fast, especially with the use of highly parallelized GPU computations.

\subsubsection{Abstract Model Simulation and Evaluation.}
With a trained MDN $g_\theta$, for an initial state $\eta_{t_0}$, the distribution of the system state at the next time instant $t + \Delta t$ is given by $g_\theta(\eta_{t_0})$, and the values of the next state $\bar{\eta}_{1}$ can be sampled from this distribution. This values then can be used to produce the next state, and so on:
\begin{equation*}
	\bar{\eta}_{i + 1} \sim g_\theta(\bar{\eta}_{i})
\end{equation*}
\noindent Every iteration in this procedure has a fixed computational cost, and therefore choosing $\Delta t$ equal to the timescale of interest, we can simulate arbitrarily long trajectories at the needed level of time resolution, without wasting computational resources.

To evaluate the abstract model, we chose a number of random initial settings and for every setting we simulate (sufficiently many) trajectories from both the original and the abstract model up to the time horizon $M \Delta t$, evaluating the distance (\ref{eq:e-close}).
Note that we train the MDN to approximate the kernel for a given $\Delta t$, i.e. it approximates model dynamics step-wise. 
However, we can evaluate the abstract model using the reward function on a time span longer than $\Delta t$. 
As noted in \cite{bortolussi_abstraction}, the idea is that a good local approximation of the kernel should result in a good global approximation on longer time scales.

%\tanja{was the discussion in previous paragraphs already in Luca's paper?}\dn{Yes, the last sentence  and the one after equation are from there, maybe we need re-phrase, but I couldn't find better words.}

\begin{example} [running example: evaluation] %\rm
For the histogram dataset,
%\tanja{you here refer to figure 2?}\dn{no, I'm speaking about dataset, fig. 2 is made of traces from it though}, 
we simulate 10000 with the stochastic simulation Gillespie algorithm (SSA) trajectories up to time 200 for 25 random initial settings, and extract state values of the discrete-time process $\tilde{\eta}_{[0, M]}$ with $\Delta t$ fixed to 20 time units.

With the trained MDN, we simulate 10000 traces starting from the same initial settings for 10 consecutive steps and therefore obtain the values $\bar{\eta}_{[0, M]}$.
\begin{figure}[t]
	\centering
	\includegraphics[width=0.3\linewidth, trim=40 40 40 40, clip]{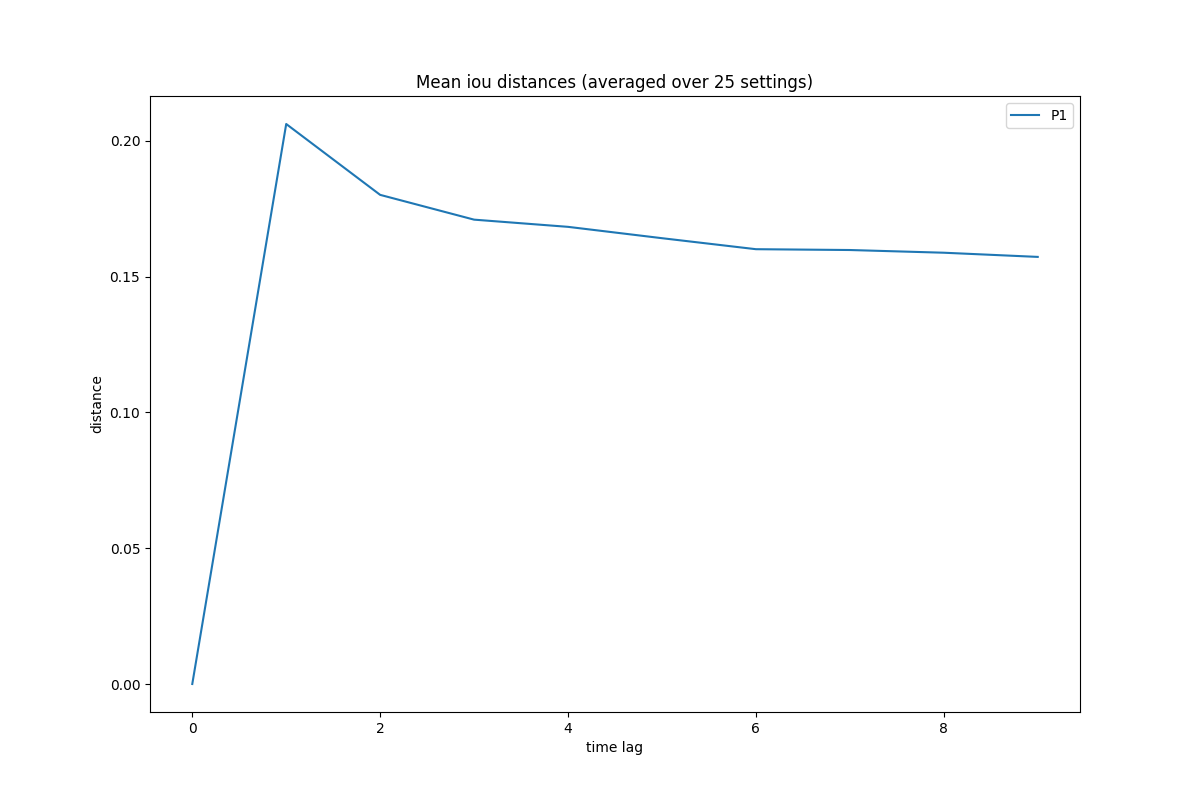}%
	\includegraphics[width=0.3\linewidth, trim=40 40 40 40, clip]{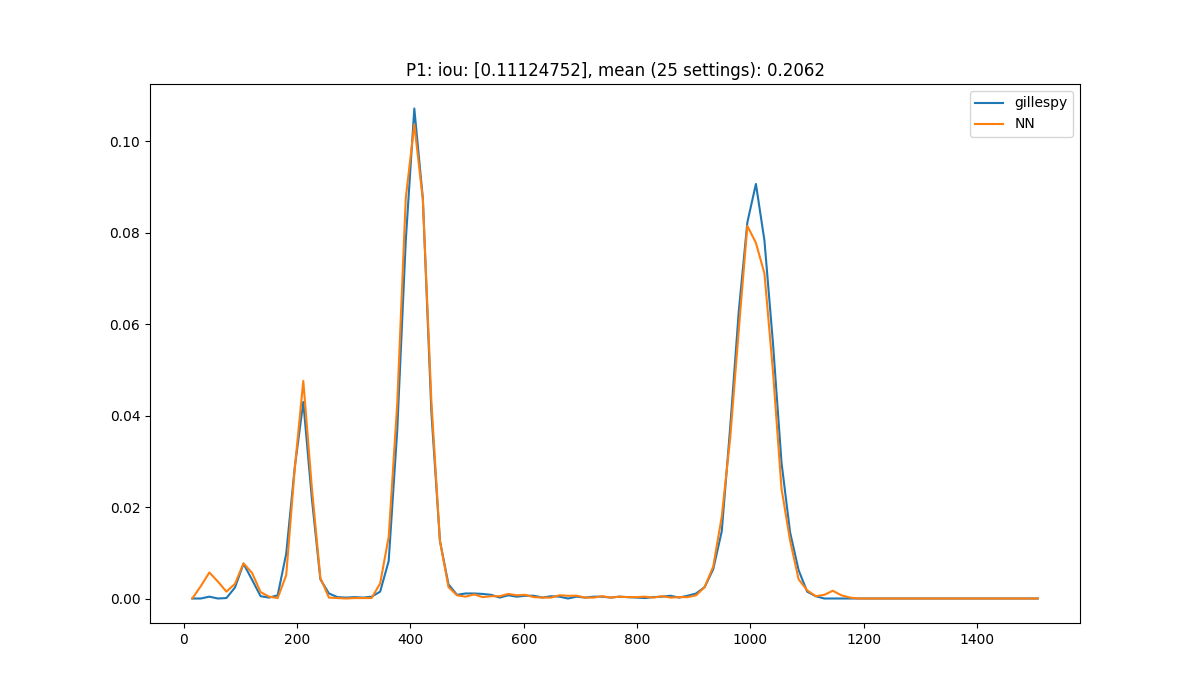}%
	\includegraphics[width=0.3\linewidth, trim=40 40 40 40, clip]{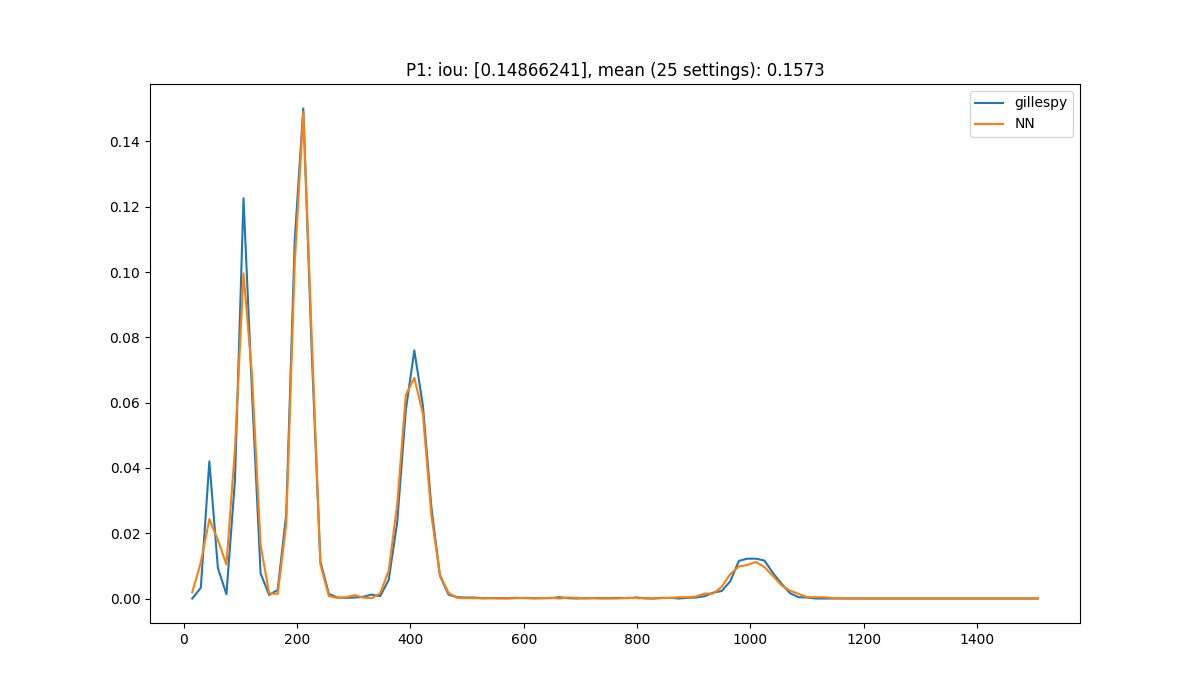}
	\caption{\footnotesize Ex.(\ref{eq:X16_}): Mean histogram distance (IoU) for different time-lags (left) and sample histograms of protein $P_1$ concentration after 1 time step (middle) and 9 time steps (right).}
	\label{fig:X16_hist_dist}
\end{figure}

Finally, we can evaluate the inequality (\ref{eq:e-close}), and estimate the average histogram distance.
For every time-step $i$ in range from 1 to $M=10$, we average the distance $d(r(\tilde{\eta}_{i}), \bar{r}(\bar{\eta}_{i}))$ over 25 initial settings. 
This displays the performance of the abstract model in predicting for many time steps in future, see Fig. \ref{fig:X16_hist_dist}. Note that, to draw the next state $\bar{\eta}_{i+1}$, the model uses its own prediction from the previous step.
\end{example}
%\begin{figure}
%	\centering
%	\includegraphics[width=0.9\linewidth]{images/X16/model2002/spec_iou.png}
%	\caption{\footnotesize X16: mean histogram distance (intersection over union) averaged over 25 different initial settings.}
%	\label{fig:X16_hist_dist}
%\end{figure}
%
%\begin{figure}
%	\centering
%	\includegraphics[width=0.9\linewidth]{images/X16/model2002/times_240.png}
%	\caption{\footnotesize X16: simulation times for MDN model (left) and Gillespie simulation (right). Times are measured to simulate traces of length 5 for different combinations of number of initial settings and number of traces for each setting.}
%\end{figure}
%
%

\section{Automated Architecture Search}
%
%\subsubsection{Neural Architecture Search.} 
The performance of machine learning algorithms depends heavily on the latent representation, i.e. a set of features. 
In contrast to simple machine learning algorithms, neural networks learn not only a mapping from representation to output but also the representation itself. 
As mentioned above, neural networks form the desired complex function from many small transformations represented by different layers. 
Each layer produces a new representation of the input features, which then can be used as an input to the following layers. 
The final representation, therefore, depends on the layer types used across the network, as well as on the graph describing connections between these layers. 

Usually, neural networks are manually engineered by the experts via the trial-and-error procedure, which is a very time-consuming and error-prone process. 
Complex tasks require models of large size, which makes model design even more challenging. 
%The evolution of models for image processing tasks is a good example of struggling to find the perfect design for effective feature extraction. 

Convolutional neural networks is a good example of a gain that comes from introducing incremental improvements in the network architecture. 
Step by step, in a series of publications, better design patterns were developed, improving the quality of models and reducing the computational demands. 
Even though a new model outperforms previous approaches, one could never argue that it is optimal. 
This raises an interest in the automated architecture search procedure that leads to the optimal model configuration given a task.

One of the first successes in this field was achieved in \cite{google_nas} where reinforcement learning was applied to discover novel architectures that outperformed human-invented models on a set of tasks such as image classification, object detection, and semantic segmentation. 
It is worth to mention that it took 2000 GPU days of training to achieve this result. %800 GPUs for three to four weeks training.
Later publications \cite{darts,proxyless_nas} introduced a gradient-based approach which significantly reduced required computational powers and allowed to achieve compatible results within one to two days using a single GPU.

In this work, we propose the algorithm inspired by \cite{darts,proxyless_nas} for the automated architecture design of MDN.
Given a dataset and only a few hyper-parameters, it learns the architecture that best fits the data.

\subsection{Our framework for automated neural network search}
Broadly speaking, all NAS methods vary within three main aspects: \emph{search space}, \emph{search policy}, and \emph{evaluation policy}.

\emph{Search space} defines which architectures can be represented in principle. 
Therefore, to define a search space we fix a set of possible operation/layer candidates, a set of rules to connect them, and the architecture size (number of connections/layers).

\emph{Search policy} describes a strategy of exploring the search space, e.g. random search, Bayesian optimization, evolutionary methods, reinforcement learning (RL), or gradient-based methods.

\emph{Evaluation policy} includes the set of metrics of interest, such as accuracy on test data, number of parameters, latency, etc.
\subsubsection{Search Space}
Similarly to DARTS architecture search method, proposed in \cite{darts}, we consider a network that consists of several computational blocks or cells. 
A cell is a directed acyclic graph consisting of an ordered sequence of $C_s$ nodes. 
Each node $x^{(i)}$ is a hidden state (latent representation) and each directed edge $(i, j)$ is associated with some operation $o^{(i,j)}$ that transforms $x^{(i)}$.

Each cell has two input nodes and a single output node. 
The input nodes are the outputs of the previous two cells (or the model inputs if there are no previous cells). 
The output node is obtained by applying an aggregating operation (e.g. \emph{sum} or \emph{mean}) on the intermediate nodes. 
Each intermediate node is computed based on all the predecessors:
\begin{equation}{\label{node}}
	x^{(j)} = \sum_{i < j} o^{(i, j)} (x^{(i)})
\end{equation}

\noindent A special zero operation is also included to indicate a lack of connection between two nodes.

\begin{figure}
	\centering
	\includegraphics[width=0.95\linewidth]{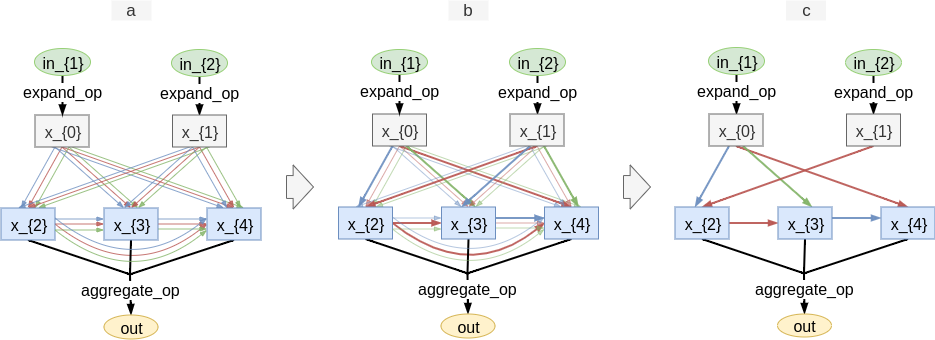}
	\caption{\footnotesize Learning a computational cell. $a):$ Operations connecting internal states are unknown and set to a mixture of candidate operations (colored edges). Every state is connected to all its predecessors. $b):$ During training, the weights for candidates are adjusted to prioritize the most important operations. $c):$ Operations with the highest weights are selected for every edge. Further, only two edges with the highest scores are selected to be inputs for each state.}
	\label{fig:nn_cell}
\end{figure}

To allow expanding the feature space within a network, we define two kinds of cells: \emph{normal cell} preserving the number of neurons(features) received at inputs, and \emph{expanding cell} that produces $d$ times more activations, where $d$ is an \emph{expansion multiplier} parameter, see Fig. \ref{fig:nn_cell} for illustration.
To serve this purpose, additional expanding operations (e.g. Dense layer) are applied to the inputs of a cell to produce the first two (input) nodes of the desired size.
The very first cell is usually an expanding cell, and the rest are normal.

Therefore, our model is defined by the number of cells $C_n$, cell size $C_s$, expansion multiplier $d$, and the set of operations on the edges. 
We consider $C_n$, $C_s$ and $d$ to be hyper-parameters defining the model backbone, and fix the set of operation candidates.
The task of architecture search thus reduces to learning the operations $o^{(i, j)}$ on the edges of each cell.
\subsubsection{Search Strategy}
The discrete search space we constructed leads to a challenging combinatorial optimisation problem, especially if we search for a model that is deep enough. 
As a neural network performs a chain of operations adjusted to each other, replacing even a single one requires a complete re-training. 
Therefore, each configuration in exponentially large search space should be trained separately.
Gradient-based methods tackle this issue by introducing a continuous relaxation for the search space so that we can leverage gradients for effective optimization.

Let $\mathcal{O} = \{ o_1, \ldots, o_N\}$ be the set of $N$ candidate operations (e.g. dense, identity, zero, etc.).
To represent any architecture in the search space, we build an over-parameterized network, where each unknown edge is set to be a mixed operation $m_{\mathcal{O}}$ with $N$ parallel paths.

First, we define weights for the edges as a softmax over $N$ real-valued architecture parameters ${\alpha_i}$ (note that outputs of softmax operation are positive and sum up to one, therefore we can treat weights $p_i$ as probabilities):
\begin{equation}
p_i = \sum_{i=1}^{N} \frac{{\rm exp}(\alpha_i)}{\sum_j {\rm exp}(\alpha_j)}.
\end{equation}

\noindent For each $m_{\mathcal{O}}$, only one operation (path) is sampled according to the probabilities $p_i$ to produce the output. Path binarization process defined in \cite{proxyless_nas} is described by:
\begin{equation}
m_{\mathcal{O}}^{Binary}(x) = \sum_{i=1}^{N} g_i o_i(x) = 
\begin{cases}
o_1(x) & \text{with probability} \ p_1, \\
\dots \\
o_N(x) & \text{with probability} \ p_N
\end{cases}
\end{equation}

\noindent where $g_i$ are binary gates:
\begin{equation}
g = \text{binarize}(p_1, ..., p_N) = 
\begin{cases}
[1, 0,\dots, 0] & \text{with probability} \ p_1, \\
\dots \\
[0, 0,\dots, 1] & \text{with probability} \ p_N.
\end{cases}
\label{eq:gates}
\end{equation}

\noindent In this way, the task of learning the architecture reduces to learning a set of parameters ${\alpha_i}$ within every cell.
The final network is obtained by replacing each mixed operation $m_{\mathcal{O}}$ by the operation $o_{i^*}$ having the largest weight: $i^*={\rm arg \ max}_i \alpha_i$.
\subsubsection{Optimization}
\begin{figure}
	\centering
	\includegraphics[width=0.95\linewidth]{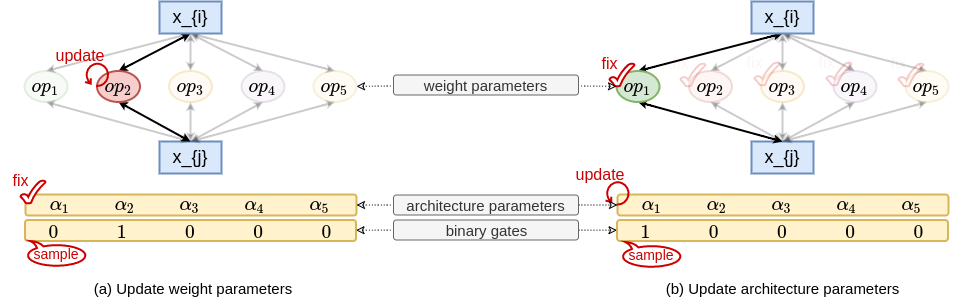}
	\caption{\footnotesize Optimization stages. \emph{left}: Weight parameters $w$ are updated on training data while $\alpha$ parameters are frozen. \emph{right}: Architecture parameters $\alpha$ are updated on validation data while $w$ parameters are frozen.}
	\label{fig:bilevel_opt}
\end{figure}

After building an over-parameterised network, our goal is to jointly optimise the architecture parameters $\alpha$ and the weights $w$ within all mixed operations. 
As discussed in \cite{darts}, the best model generalisation is achieved by reformulating our objective as a bi-level optimisation problem. 
We minimise the validation loss $\mathcal{L}_{val}(w^*, \alpha^*)$ w.r.t. $\alpha ^*$, where the weights $w^*$ are obtained by minimising the training loss $\mathcal{L}_{train}(w, \alpha^*)$. 
In other words, training is performed by altering two separate stages for several epochs each, see Fig. \ref{fig:bilevel_opt}.

When training weight parameters, we first freeze the architecture parameters $\alpha$. 
Then for every example in the training dataset, we sample binary gates according to (\ref{eq:gates}) and update the weight parameters of active paths via standard gradient descent.

When training architecture parameters, we freeze the weight parameters and update the architecture parameters on validation data. For every batch, binary gates are sampled w.r.t updated architecture parameters.

However, due to the nature of the binarization procedure, the paths probabilities $p_i$ are not directly involved in the computational graph, which means that we can not directly compute gradients 
\begin{equation}
	\frac{\partial L}{\partial \alpha_i} = \sum_{j=1}^{N} \frac{\partial L}{\partial p_j} \frac{\partial p_j}{\partial \alpha_i}
	\label{eq:gradients}
\end{equation}

\noindent to update $\alpha_i$ using the gradient descent.
As it was proposed in \cite{proxyless_nas,binaryconnect}, we update the architecture parameters using the gradient w.r.t. its corresponding binary gate $g_i$, i.e. using $\partial L/ \partial g_i$ instead of $\partial L/ \partial p_i$ in (\ref{eq:gradients}).

\begin{example} [running example: architecture search] %\rm
We search for the network consisting of 2 cells each of size 2, the first cell is an expanding cell. We train for 100 epochs in total: first 20 epochs only networks weights are updated, and the following 80 epochs training is performed as on Fig. \ref{fig:bilevel_opt}.
See Fig. \ref{fig:X16_nn} for the example of learned architecture.
\begin{figure}[t]
	\centering
	\includegraphics[width=0.9\linewidth]{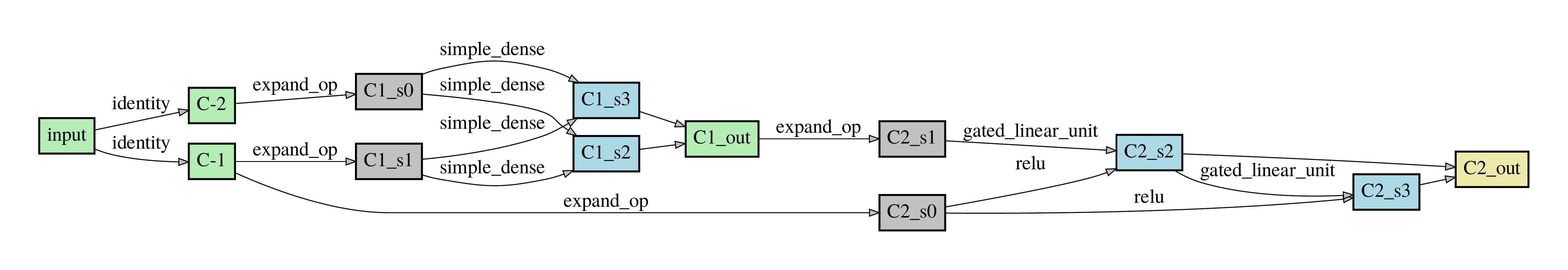}
	\caption{\footnotesize Ex.(\ref{eq:X16_}): learned network structure. Gray rectangles represent input nodes, blue - intermediate nodes, Green and yellow - output nodes (or cell inputs). Intermediate nodes compute the sum of values on incoming edges, output nodes - the average.}
	\label{fig:X16_nn}
\end{figure}
\end{example}
\section{Implementation}
%\noindent We implemented the proposed abstraction framework as a python library StochNetV2 available on GitHub at \url{https://github.com/dennerepin/StochNetV2}.
The library has implementations for all parts of the workflow:
\begin{itemize}
	\item defining a custom CRN model or importing SBML/Kappa file,
	\item producing CRN trajectories and creating datasets for training and evaluation,
	\item training custom models and automated architecture search,
	\item model evaluation,
	\item generating traces with trained model,
	\item various visualisations (traces, histograms, mixture parameters, architecture design, model benchmarking etc.).
\end{itemize}

\noindent To simulate traces more effectively, we provide scripts that run many simulations in parallel using multi-threading routines.

We use \emph{luigi} \cite{luigi} package as a workflow manager, which allows creating complex pipelines and managing complex dependencies.
Neural networks, random variables, and optimisation algorithms are implemented in \emph{TensorFlow} \cite{tensorflow} and deep learning framework.
\subsubsection{CRN Models and Simulation Data.}
CRN models are handled by \emph{gillespy} python library \cite{gillespy}, which allows to define a custom class for model or import a model in SBML format. Note that not all models can be imported correctly, due to high variability of the SBML format. In those cases one can use a custom class with some pre-processing of the imported model. 
%
%We also support rule-based models defined in the Kappa language \cite{kappy} with the use of the \emph{kappy} python library.

Simulated trajectories we split into training examples $(x, y) := (\tilde{\eta}_{i}, \tilde{\eta}_{i+1})$. As neural networks show better convergence when the training data is standardised so that it varies in a reasonable range (e.g. [-1, 1] or [0, 1]) or has zero mean and variance, we apply a preprocess step to the input data such that it is scaled to a desired range. 
Then it is split into training (80\%) and validation (20\%) parts. 
The training dataset is used to optimise network weights, and validation dataset is used to optimise architecture parameters.

To increase generalisation capabilities of the model, it is important to build the dataset that covers the most variability of the original process.
Although having more training data is always beneficial for a model, it increases training time.
Therefore, depending on the variation of trajectories starting from the same initial conditions, we might prefer to run a few simulations for many initial conditions or more simulations for fewer initial conditions.
When generating the dataset for evaluation, to make histograms more consistent, we usually simulate much more trajectories (from 1000 to 10000) for several initial settings.

\subsubsection{Network Structure and Computational Cells.}
In our experiments, we learn the network typically constructed from two to three computational cells each of size 2 to 4. The first cell is expanding with a multiplier in a range from 10 to 20, and other cells are normal cells.
Having multiple cells is not necessary, so it may consist of only one (larger) cell.

A computational cell described in the previous sections may also be altered. 
First, the \emph{expanding operations} in the beginning of a cell can be represented either by Dense layers or by identity operations with tiling. For instance, for a multiplier 2 it transforms a vector $(x_1, x_2,...,x_n)$ into a vector $(x_1,x_1, x_2, x_2,...,x_n, x_n)$. 
Second, we can vary the number of intermediate nodes of a cell being aggregated to produce the output (e.g. all intermediate nodes or the last one, two, etc.), as well as the aggregating operation itself (e.g. $sum$ or $mean$). Smaller number of aggregated nodes may lead to smaller cells after removing redundant connections at the final stage of architecture selection. If all edges connecting some of the intermediate nodes with the output are pruned. 

%Additionally, we implemented an alternative idea of mixed operation, which is originally proposed in \cite{darts}, where the outputs of candidate operations are not sampled with probabilities $p_i$, but rather added together and multiplied with the corresponding values $p_i$. However, this approach introduces significant discrepancies to layer activations after the final architecture is selected (as we replace a weighted sum of many operations with a single operation), and therefore the model has to be re-trained. The approach from \cite{proxyless_nas}, described in the NAS section doesn't have this issue, because in this case we just stop sampling operations and fix the choice on one of them. For this reason, we prefer the latter approach with path sampling.

\subsubsection{Random Variables and MDN Training.}
Our implementation has various random variables from Gaussian family: \emph{Normal Diagonal}, \emph{Normal Triangular}, \emph{Log-Normal Diagonal}.
Different combinations of these variables can be selected as the components for the mixture distribution, as long as their samples have the same dimensionality.
In our experiments, we usually use from 5 to 8 components of Normal Diagonal variables. Replacing 2-3 Normal Diagonal components with the same number of Normal Triangular components sometimes improves model quality, though slows down both training and simulation times.

When training the architecture search, we have three main stages: 
\begin{itemize}
	\item \emph{heat-up stage}, when weight parameters are trained for 10-20 epochs without updating the architecture parameters,
	\item \emph{architecture search stage}, when weight parameters and architecture parameters are updated as described in NAS section (see Fig. \ref{fig:bilevel_opt}), for 50 to 100 epochs in total, each turn of updating weight/architecture parameters typically lasts 3 to 5 epochs,
	\item \emph{fine-tuning stage}, when the final architecture is fine-tuned for 10 to 20 epochs.
\end{itemize}

\section{Conclusions}
%%\begin{table}[t]
%	\begin{tabular}{| l | c | c | c | c | c |}
%		\hline
%		{\bf Task}                            & \multicolumn{5}{ |c| }{{\bf Time}}  \\ \hline
%		&  {\bf EGFR}   & {\bf Gene} & {\bf X16}  & {\bf X40} &  {\bf X44}         \\ \hline
%		Generate training data                &    820.8 s.   &  999.5 s.  &  198.7 s.  &  13.1 s.  &  468.3 s.          \\ \hline
%		Format dataset                        &    1.32 s.    &  0.72 s.   &  0.66 s.   &  0.17 s.  &  0.74 s.           \\ \hline
%		Train model                           &    213 min.   &  25 min.   &  49 min.   &  24 min.  &  81 min.           \\ \hline
%		{\bf Generate histogram data (SSA)}   &    46.8 s.    &  9825.4 s. &  1274.5 s. &  317.9 s. &  3048.2 s.         \\ \hline
%		{\bf Generate histogram data (MDN)}   &    41.3 s.    &  19.4 s.   &  37.7      &  43.1 s.  &  41.9 s.           \\ \hline
%	\end{tabular}
%	\caption{\footnotesize
%		Execution time required to complete each step of the abstraction pipeline for different models. The last two rows display the difference in simulation times between the Gillespie SSA algorithm and the MDN abstract model. All simulations here performed in similar conditions: for every model we simulate the same number of trajectories up to the same time horizon (typically 10-20 time steps $\Delta t$), using the same number of CPU cores.
%	}
%	\label{tbl:res_speed}
%\end{table}

\begin{centering}
\begin{table}[t]
\centering
	\begin{tabular}{| l | c | c | c | }
		\hline
		{\bf Task}                            & \multicolumn{3}{ |c| }{{\bf Time}}  \\ \hline
		&\;\;\;\; \;\;{\bf EGFR} \;\;\;\;\;\;  &\;\;\;\;\;\; {\bf Gene}\;\;\;\;\;\;\ &\;\;\;\;\;\;\ {\bf X16} \;\;\;\;\;\;\
		\\ 
		\hline
		Generate training data                &    820.8 s.   &  999.5 s.  &  198.7 s.       \\ \hline
		Format dataset                        &    1.32 s.    &  0.72 s.   &  0.66 s.    \\ \hline
		Train model                           &    213 min.   &  25 min.   &  49 min.         \\ \hline
		{\bf Generate histogram data (SSA)}   &    46.8 s.    &  9825.4 s. &  1274.5 s.        \\ \hline
		{\bf Generate histogram data (MDN)}   &    41.3 s.    &  19.4 s.   &  37.7 s.             \\ \hline
	\end{tabular}
	\vspace{2mm}
	\caption{\footnotesize
		Execution time required to complete each step of the abstraction pipeline for different models. The last two rows display the difference in simulation times between the Gillespie SSA algorithm and the MDN abstract model. All simulations here performed in similar conditions: for every model we simulate the same number of trajectories up to the same time horizon (typically 10-20 time steps $\Delta t$), using the same number of CPU cores.
	}
	\label{tbl:res_speed}
\end{table}
\end{centering}

In this paper, we proposed how to automatise deep abstractions for stochastic CRNs, 
through learning the optimal neural network architecture along with learning the transition kernel of the abstract process. 
Automated search of the architecture makes the method applicable directly to any given CRN, 
which is time-saving for deep learning experts and crucial for non-specialists. 
Contrary to the manual approach where the user has to create a neural network by hand, test it for his use-case, and adopt it accordingly, our method allows to find a solution with minimal efforts within a reasonable amount of time.
We implement the method and demonstrated its performance on three representative CRNs, two of which exhibit multi-modal emergent phenotypes. 
Compared to the plain stochastic simulation, our method is significantly faster in almost all use-cases, see Table \ref{tbl:res_speed}. 

The proposed methodology, especially automated architecture search, enables fast simulation of computationally expensive Markov processes. 
As such, it opens up possibilities for efficiently simulating interactions between many individual entities, each described by a complex reaction network.
Although our method is generic with respect to the model, it is sensitive to model population size. 
In short-term future work, we would like to relax this limitation and develop a strategy that is agnostic to the size of the system being modelled.

\bibliographystyle{splncs04}
\bibliography{bibliography}
%
%\clearpage
\section{Supplementary Material}
\subsection{Background and Preliminaries}

%\begin{figure}[]
%	\centering
%	\includegraphics[width=0.9\linewidth]{images/MDN_scheme.eps}
%	\caption{\footnotesize Mixture Density Network structure. Given ${\bf x}$, neural network outputs values $ \mu_i $ and $ \sigma_i , i=1,...,m$ that define $m$ Gaussian distributions. Weighted by mixing coefficients $\alpha_i$, they form a mixture density - conditional probability density $p({\bf y}|{\bf x})$.}
%	\label{fig:mdn}
%\end{figure}

\subsection{EGFR}

Epidermal growth-factor receptor (EGFR) reaction model of cellular signal transduction, with 25 reactions and 23 different molecular species:

\begin{subequations}\label{eq:EGFR}
	\begin{align*}
	& R + EGF          \xrightleftharpoons[k_{1b}]{k_{1f}}                        Ra  \\[-1.5ex]
	& 2 Ra             \xrightleftharpoons[k_{2b}]{k_{2f}}                        R2  \\[-1.5ex]
	& R2               \xrightleftharpoons[k_{3b}]{k_{3f}}                        RP  \\[-1.5ex]
	& RP               \xrightarrow{v_{4} * RP / (k_{4} + RP)}                    R2  \\[-1.5ex]
	& RP + PLCg        \xrightleftharpoons[k_{5b}]{k_{5f}}                     PRLCg  \\[-1.5ex]
	& RPLCg            \xrightleftharpoons[k_{6b}]{k_{6f}}                    RPLCgP  \\[-1.5ex]
	& RPLCgP           \xrightleftharpoons[k_{7b}]{k_{7f}}                PLCgP + RP  \\[-1.5ex] %\tag{\ref{eq:EGFR}}
	& PLCgP            \xrightarrow{v_{8} * PLCgP / (k_{8} + PLCgP)}            PLCg  \\[-1.5ex]
	& Grb + RP         \xrightleftharpoons[k_{9b}]{k_{9f}}                        RG  \\[-1.5ex]
	& RG + SOS         \xrightleftharpoons[k_{10b}]{k_{10f}}                     RGS  \\[-1.5ex]
	& RGS              \xrightleftharpoons[k_{11b}]{k_{11f}}                 GS + RP  \\[-1.5ex]
	& GS               \xrightleftharpoons[k_{12b}]{k_{12f}}               Grb + SOS  \\[-1.5ex]
	& Shc + RP         \xrightleftharpoons[k_{13b}]{k_{13f}}                     RSh  \\[-1.5ex] \tag{\ref{eq:EGFR}}
	& RSh              \xrightleftharpoons[k_{14b}]{k_{14f}}                    RShP  \\[-1.5ex]
	& RShP             \xrightleftharpoons[k_{15b}]{k_{15f}}                RP + ShP  \\[-1.5ex]
	& ShP              \xrightarrow{v_{16} * ShP / (k_{16} + ShP)}               Shc  \\[-1.5ex]
	& RShP + Grb       \xrightleftharpoons[k_{17b}]{k_{17f}}                    RShG  \\[-1.5ex]
	& RShG             \xrightleftharpoons[k_{18b}]{k_{18f}}                ShG + RP  \\[-1.5ex]
	& SOS + RShG       \xrightleftharpoons[k_{19b}]{k_{19f}}                   RShGS  \\[-1.5ex] %\tag{\ref{eq:EGFR}}
	& RShGS            \xrightleftharpoons[k_{20b}]{k_{20f}}               ShGS + RP  \\[-1.5ex]
	& Grb + ShP        \xrightleftharpoons[k_{21b}]{k_{21f}}                    ShG   \\[-1.5ex]
	& ShG + SOS        \xrightleftharpoons[k_{22b}]{k_{22f}}                   ShGS   \\[-1.5ex]
	& ShGS             \xrightleftharpoons[k_{23b}]{k_{23f}}                GS + ShP  \\[-1.5ex]
	& RShP + GS        \xrightleftharpoons[k_{24b}]{k_{24f}}                   RShGS  \\[-1.5ex]
	& PLCgP            \xrightleftharpoons[k_{25b}]{k_{25f}}                   PLCgl 
	\end{align*}
\end{subequations}

\begin{figure}
	\centering
	\includegraphics[width=0.9\linewidth]{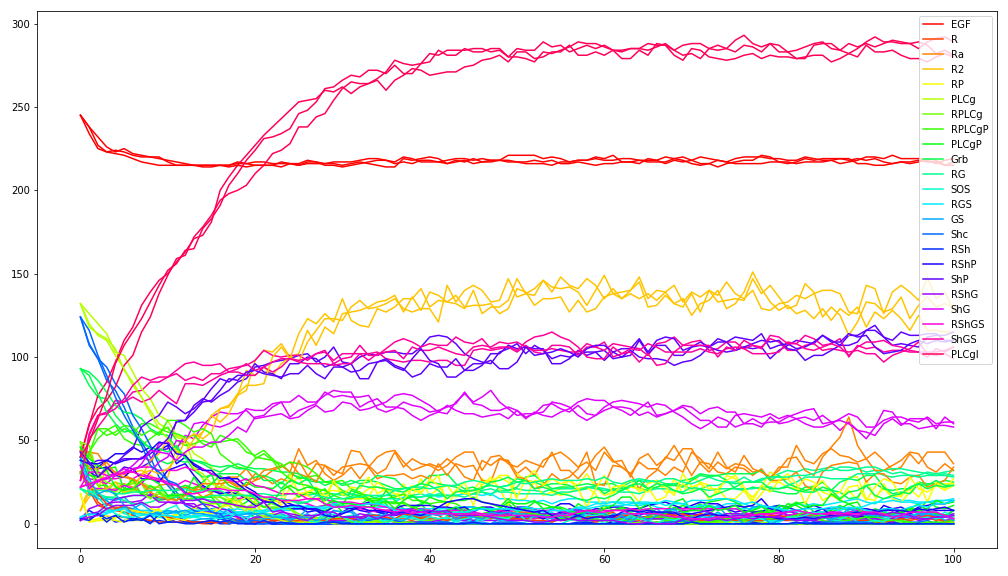}
	\caption{\footnotesize Three sample trajectories of EGFR network starting from same initial state for 50 time steps $\Delta t =0.5$}
\end{figure}

\begin{figure}
	\centering
	\includegraphics[width=0.48\linewidth, trim=80 60 60 80, clip]{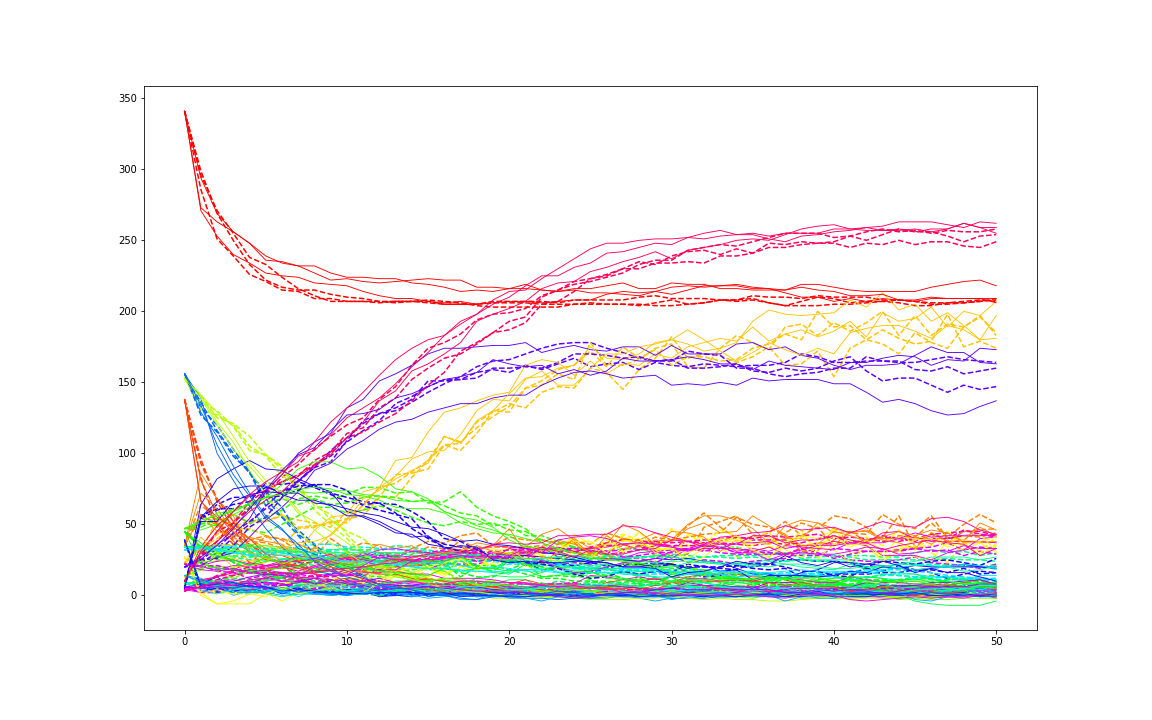}%
	\includegraphics[width=0.48\linewidth, trim=80 60 60 80, clip]{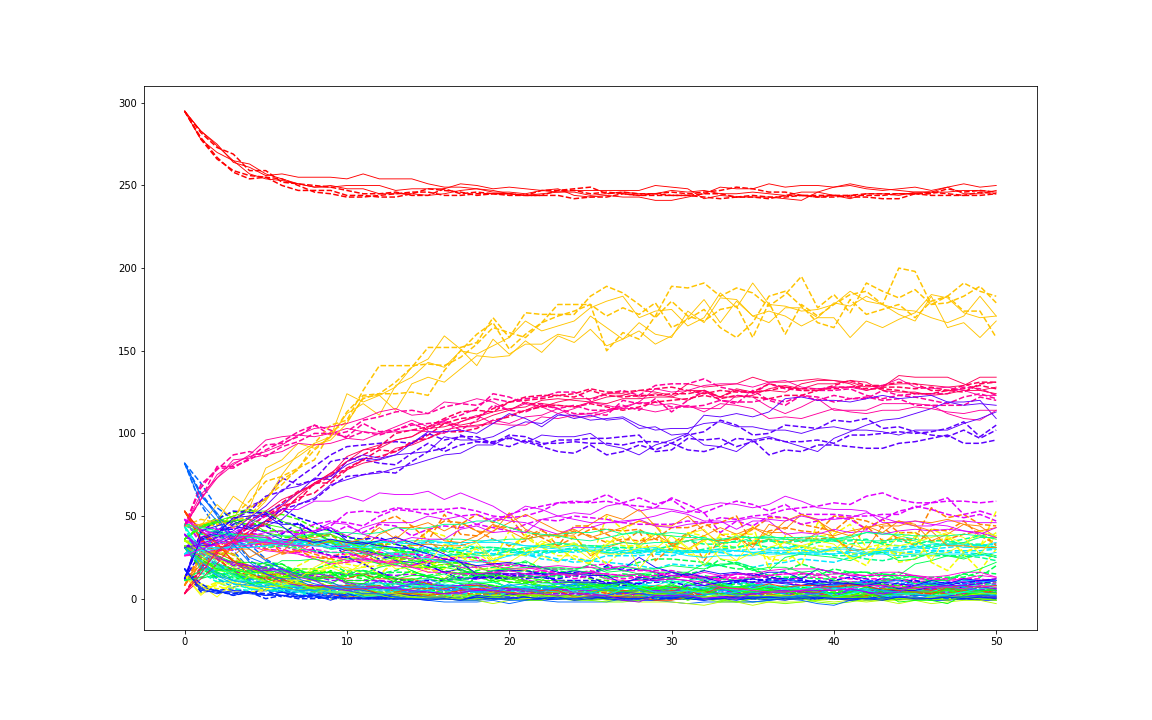}
	\caption{\footnotesize EGFR: traces simulated by Gillespie algorithm (dashed lines) and MDN (full lines) for 50 consecutive time steps, $\Delta t =0.5$.}
\end{figure}

\begin{figure}
	\centering
	\includegraphics[width=0.9\linewidth]{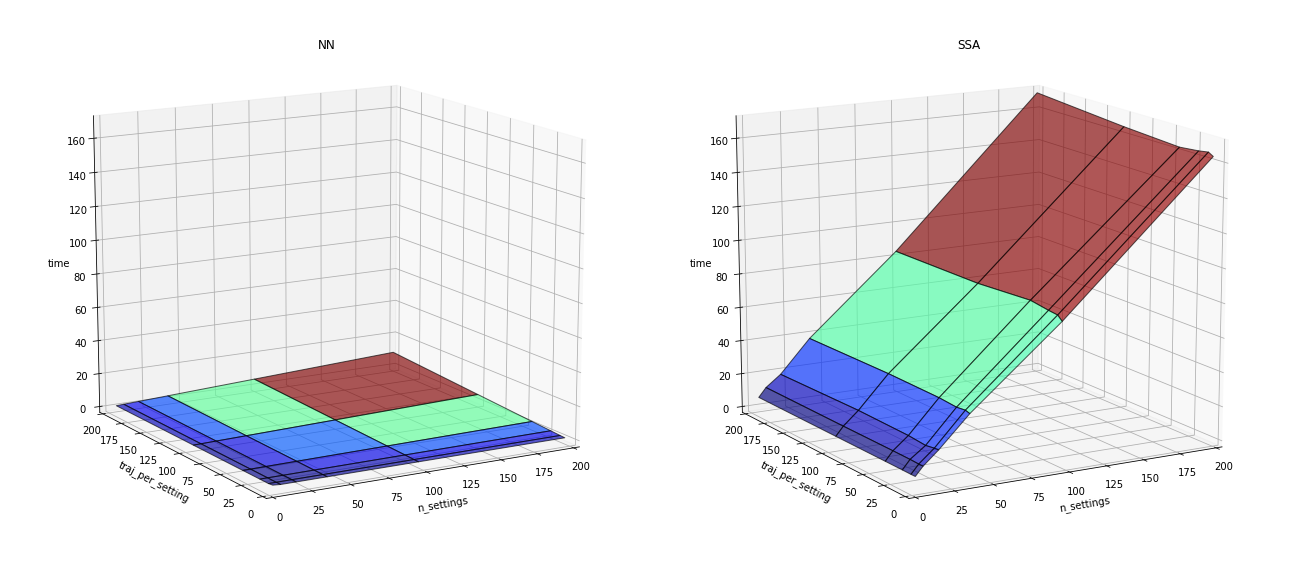}
	\caption{\footnotesize Simulation times for EGFR model. Left: MDN model, right: Gillespie simulation. Times are measured to simulate traces of length 5 for different combinations of number of initial settings and number of traces for each setting.}
\end{figure}

\begin{figure}
	\centering
	\includegraphics[width=0.9\linewidth]{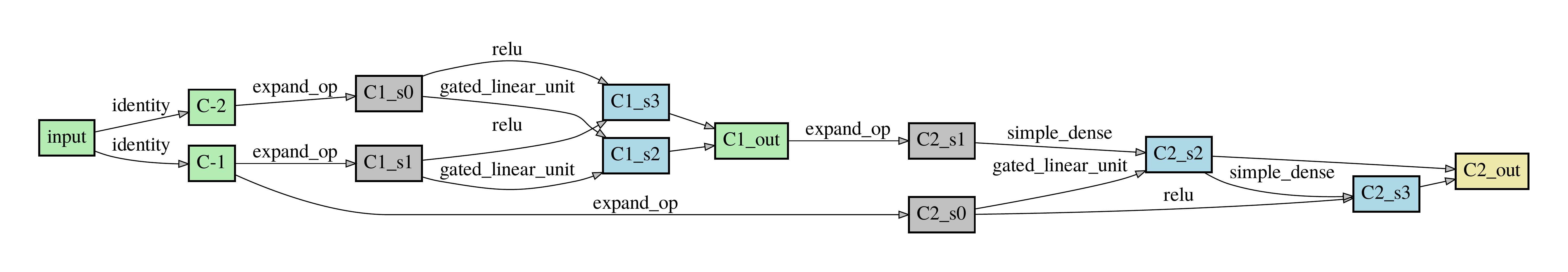}
	\caption{\footnotesize EGFR: learned network structure. Gray rectangles represent input nodes, blue - intermediate nodes, Green and yellow - output nodes (or cell inputs). Intermediate nodes compute the sum of values on incoming edges, output nodes - the average.}
\end{figure}

\subsection{Gene}

Self-regulated gene network \cite{gene_model,bortolussi_abstraction}: a single gene G is transcribed to produce copies of a mRNA signal molecule M, which are in turn translated into copies of a protein P; P acts as a repressor with respect to G - it binds to a DNA-silencer region, inhibiting gene transcription. 

\begin{equation}
\begin{split}
& G     \xrightarrow{k_{prodM}}     G + M     \\
& M     \xrightarrow{k_{prodP}}     M + P     \\
& M     \xrightarrow{k_{degM}}      \emptyset \\
& P     \xrightarrow{k_{degP}}      \emptyset \\
& G + P \xrightarrow{k_{bindP}}      Gb       \\
& Gb    \xrightarrow{k_{unbindP}}   G + P
\end{split}
\end{equation}

\begin{figure}
	\centering
	\includegraphics[width=0.9\linewidth]{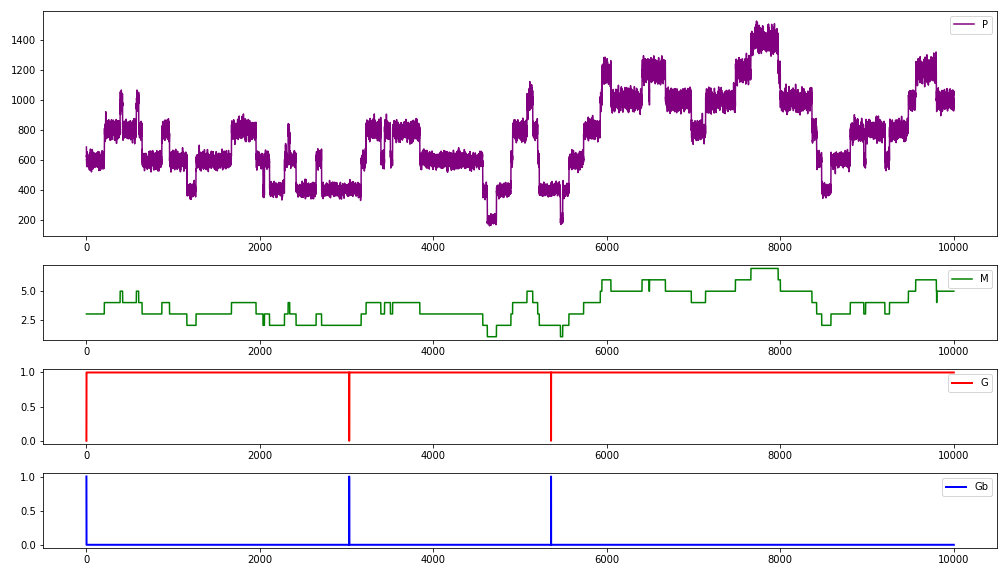}
	\caption{\footnotesize Sample trajectory of gene regulatory network.}
\end{figure}

\begin{figure}
	\centering
	\includegraphics[width=0.9\linewidth]{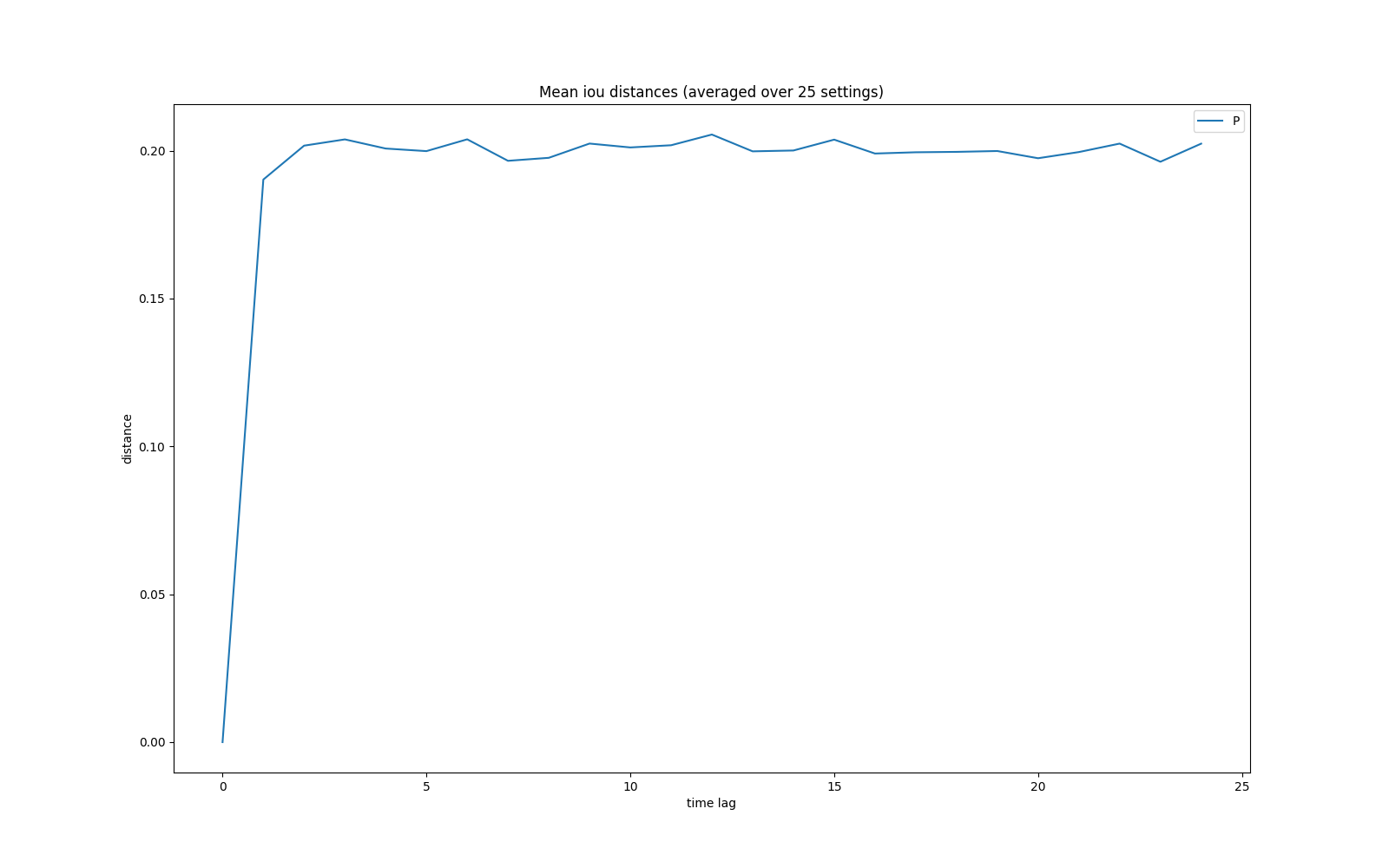}
	\caption{\footnotesize Gene: mean histogram distance (intersection over union) averaged over 25 different initial settings.}
\end{figure}

\begin{figure}
	\centering
	\includegraphics[width=0.45\linewidth, trim=40 40 40 40, clip]{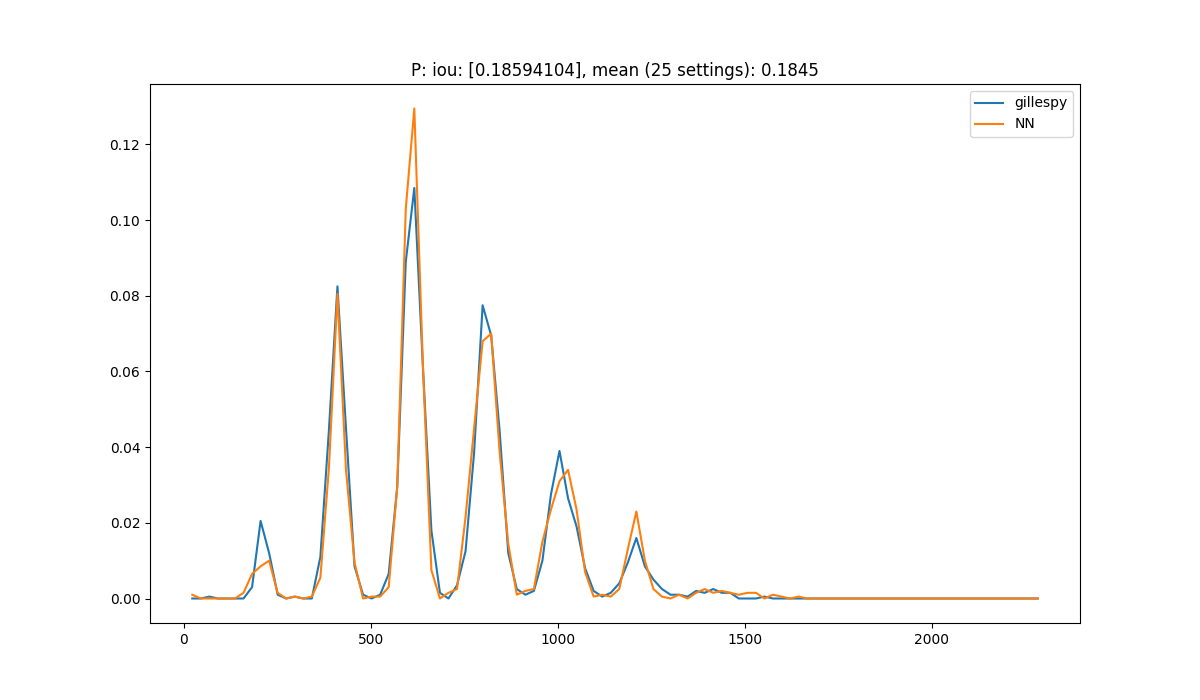}%
	\includegraphics[width=0.45\linewidth, trim=40 40 40 40, clip]{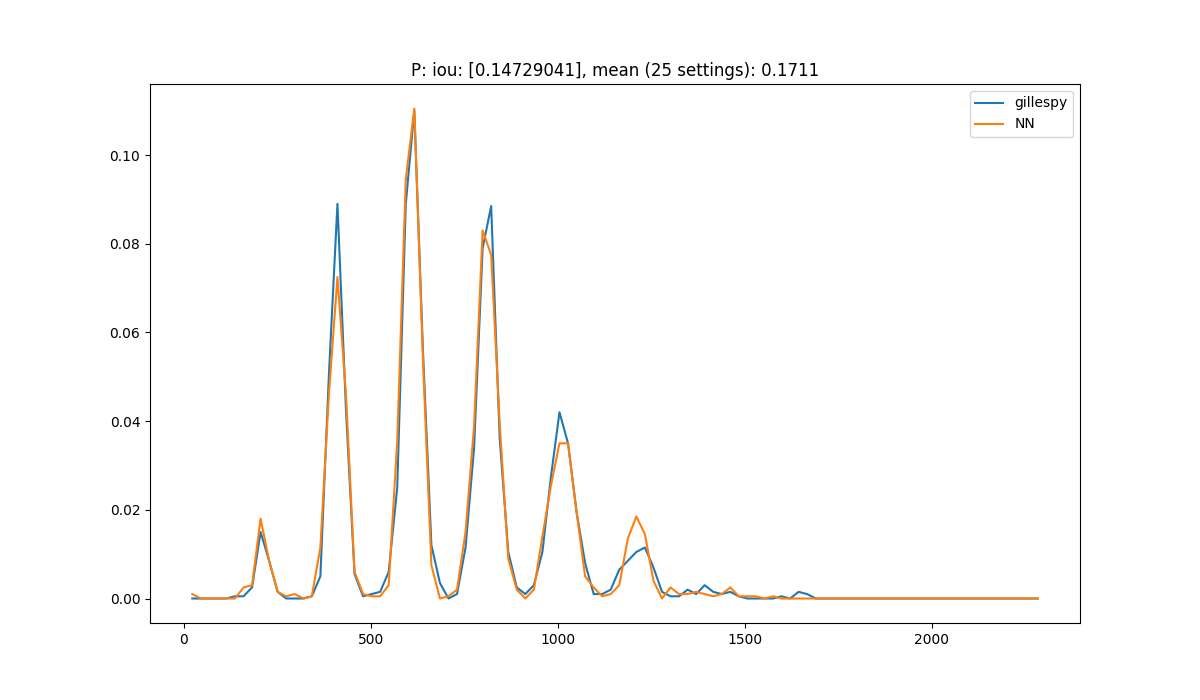}
	\caption{\footnotesize Gene: histograms of protein $P$ concentration after 5 time steps (left) and 25 time steps (right).}
\end{figure}

\begin{figure}
	\centering
	\includegraphics[width=0.9\linewidth]{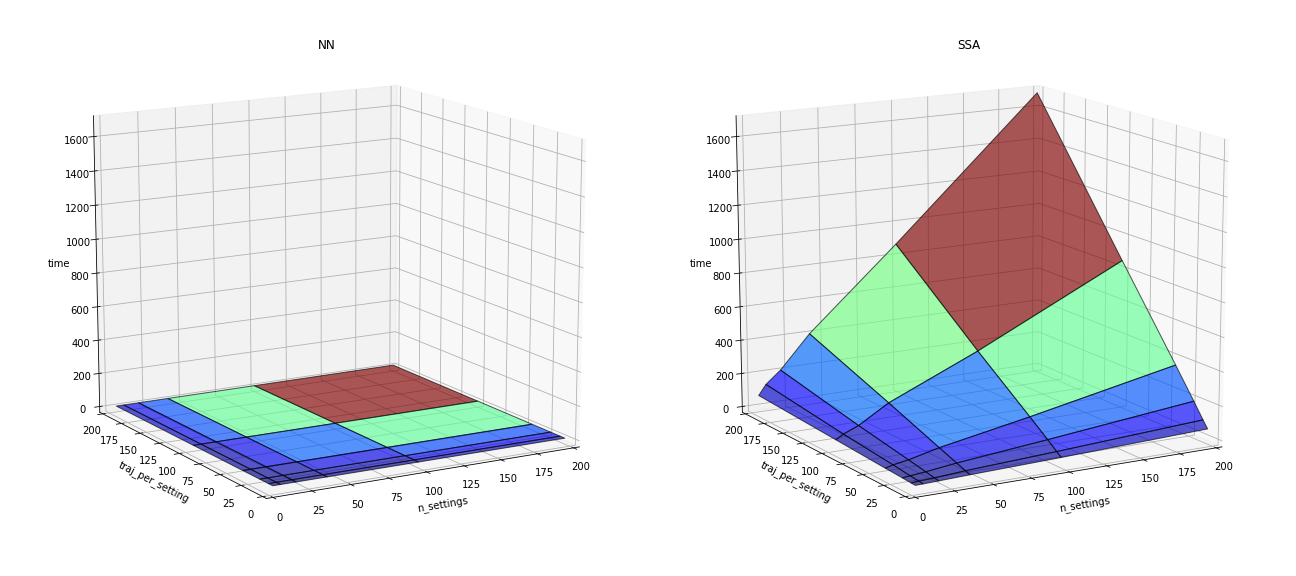}
	\caption{\footnotesize Gene: simulation times for MDN model (left) and Gillespie simulation (right). Times are measured to simulate traces of length 5 for different combinations of number of initial settings and number of traces for each setting.}
\end{figure}

\begin{figure}
	\centering
	\includegraphics[width=0.9\linewidth]{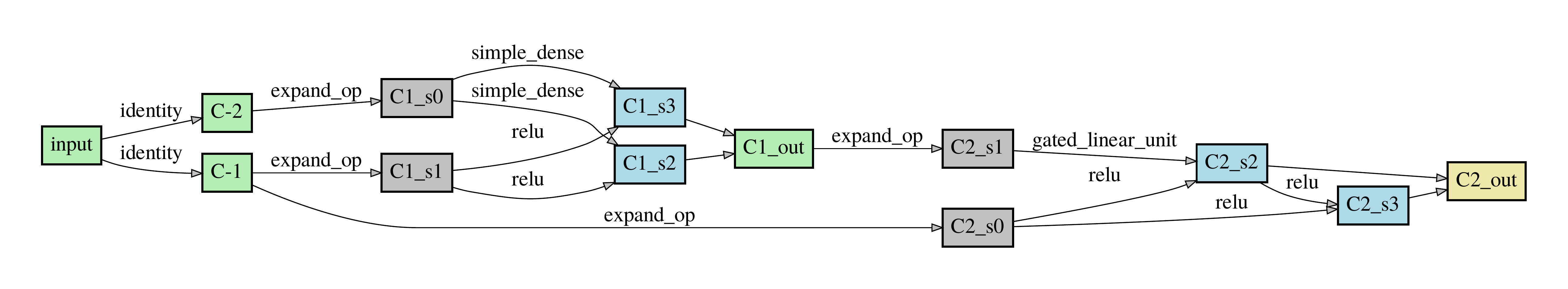}
	\caption{\footnotesize Gene: learned network structure. Gray rectangles represent input nodes, blue - intermediate nodes, Green and yellow - output nodes (or cell inputs). Intermediate nodes compute the sum of values on incoming edges, output nodes - the average.}
\end{figure}

\subsection{X16}

The following fast-slow network (\cite{crn_models}) displays interesting dynamics with multimodal species distribution changing through time, as well as for different initial settings:

\begin{equation}
\begin{split}
& G_1   \xrightleftharpoons[\alpha_{12}]{\alpha_{11}}                     G_1 + P_1  \\
& G_2   \xrightarrow{\alpha_{21}}                                         G_2 + P_1  \\
& P_1   \xrightarrow{\beta_{1}}                                           \emptyset  \\
& G_1   \xrightleftharpoons[\epsilon \gamma_{21}]{\epsilon \gamma_{12}}   G_2, \ \ 0 < \epsilon \ll 1
\end{split}
\label{eq:X16}
\end{equation}
Network (\ref{eq:X16}) may be interpreted as describing a gene slowly switching between two expressions $\textrm{G}_1$ and $\textrm{G}_2$. 
When in state $\textrm{G}_1$, the gene produces and degrades protein $\textrm{P}_1$, while when in state $\textrm{G}_1$, it only produces $\textrm{P}_1$, but generally at a different rate than when it is in state $\textrm{G}_1$. 
Furthermore, $\textrm{P}_1$ may also spontaneously degrade.

\begin{figure}
	\centering
	\includegraphics[width=0.9\linewidth]{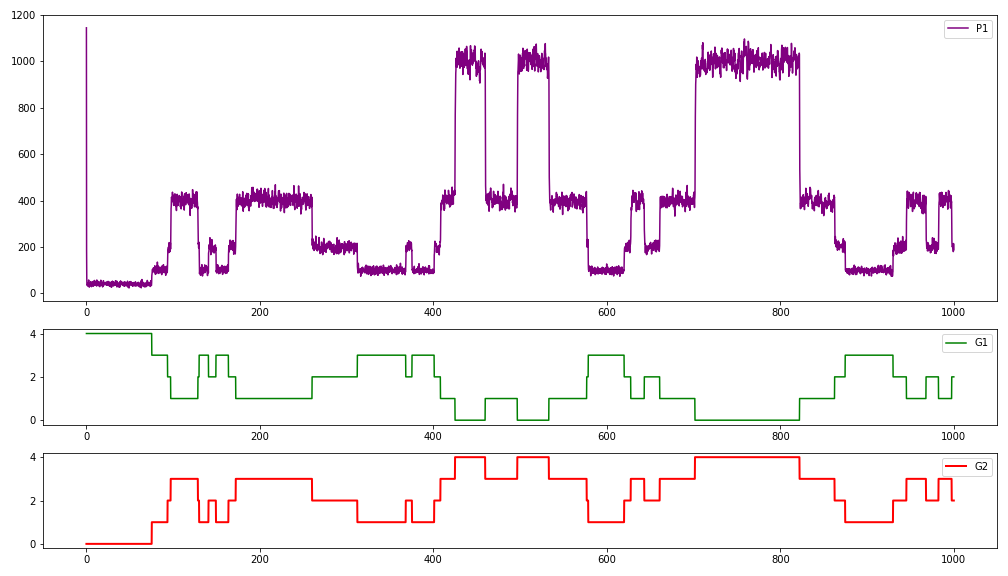}
	\caption{\footnotesize Sample trajectory of X16 network.}
\end{figure}

%\begin{figure}
%	\centering
%	\includegraphics[width=0.45\linewidth]{images/X16/X16_hist_0_time200.png}%
%	\includegraphics[width=0.45\linewidth]{images/X16/X16_hist_2_time200.png}
%	\caption{\footnotesize Steady state distribution of protein $P_1$ (histogram at time 200) for different initial settings, X16 network.}
%\end{figure}

\begin{figure}
	\centering
	\includegraphics[width=0.9\linewidth]{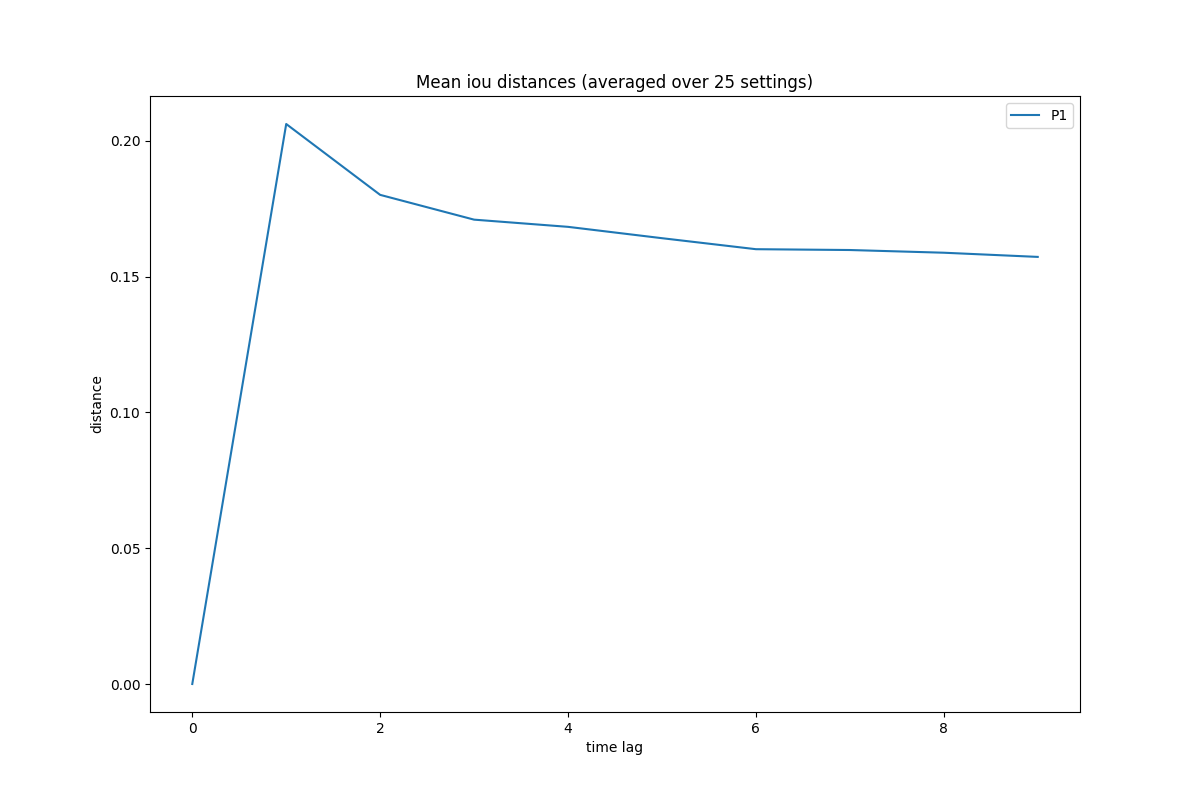}
	\caption{\footnotesize X16: mean histogram distance (intersection over union) averaged over 25 different initial settings.}
\end{figure}

\begin{figure}
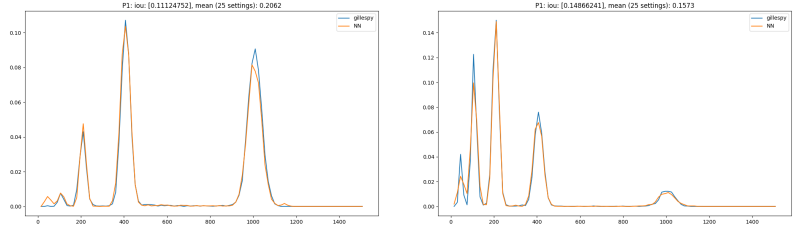

	\centering
	\includegraphics[width=0.45\linewidth, trim=40 40 40 40, clip]{images/X16/model2002/P1_t1_s1.png}%
	\includegraphics[width=0.45\linewidth, trim=40 40 40 40, clip]{images/X16/model2002/P1_t9_s1.png}
	\caption{\footnotesize X16: histograms of protein $P_1$ concentration after 1 time step (left) and 9 time steps (right).}
\end{figure}

\begin{figure}
	\centering
	\includegraphics[width=0.9\linewidth]{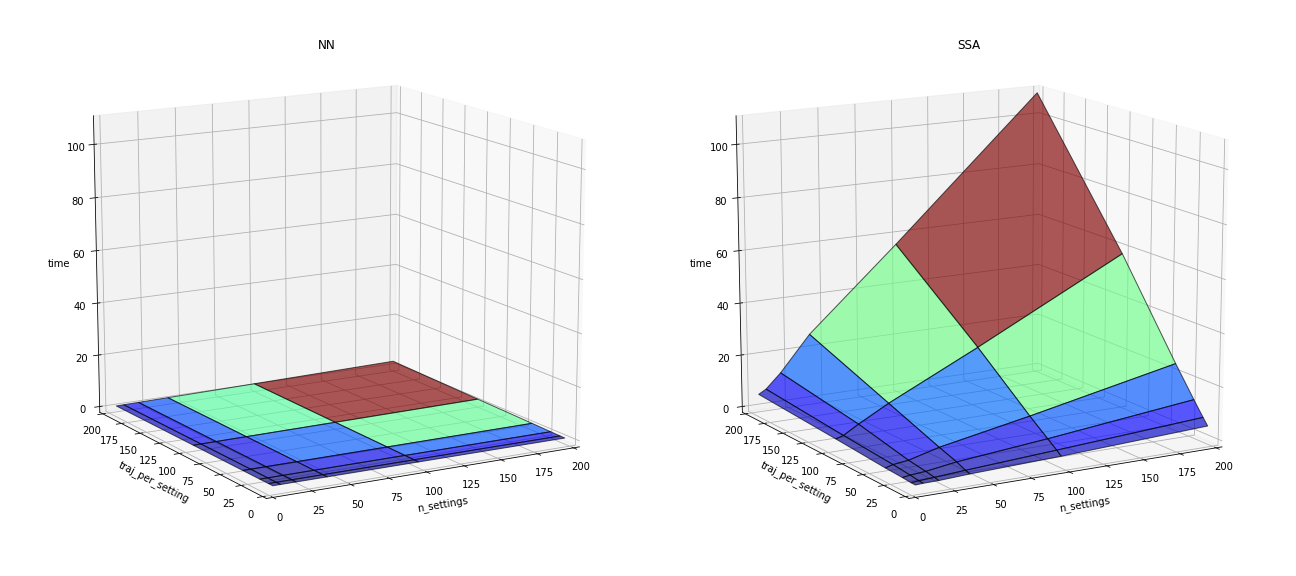}
	\caption{\footnotesize X16: simulation times for MDN model (left) and Gillespie simulation (right). Times are measured to simulate traces of length 5 for different combinations of number of initial settings and number of traces for each setting.}
\end{figure}

\begin{figure}
	\centering
	\includegraphics[width=0.9\linewidth]{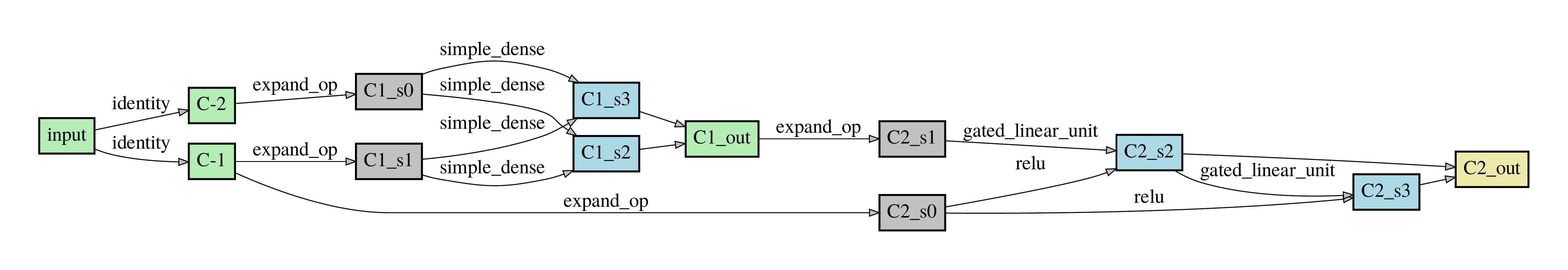}
	\caption{\footnotesize X16: learned network structure. Gray rectangles represent input nodes, blue - intermediate nodes, Green and yellow - output nodes (or cell inputs). Intermediate nodes compute the sum of values on incoming edges, output nodes - the average.}
\end{figure}

\end{document}